\documentclass[12pt]{article}
\usepackage{latexsym,amsmath,amssymb,}
\usepackage{graphicx}
\usepackage{ulem}

\begin{document}

\title{Quasi-normal modes,  bifurcations and non-uniqueness of charged scalar-tensor black holes}

\author{Daniela D. Doneva$^{1,2}$\thanks{E-mail: ddoneva@phys.uni-sofia.bg}\,,\:  Stoytcho S. Yazadjiev$^{3}$ \thanks{E-mail:
yazad@phys.uni-sofia.bg}
\,,\: Kostas D. Kokkotas$^{2,4}$ \thanks{E-mail: kostas.kokkotas@uni-tuebingen.de} \,\,\\{\footnotesize  ${}^{1}$Dept.
of Astronomy,
                Faculty of Physics, St.Kliment Ohridski University of Sofia}\\
                {\footnotesize  5, James Bourchier Blvd., 1164 Sofia, Bulgaria }\\\\[-3.mm]
      {  \footnotesize ${}^{2}$ Theoretical Astrophysics, Eberhard-Karls University of T\"ubingen, T\"ubingen 72076, Germany }\\
      \\
      {  \footnotesize ${}^{3}$Dept. of Theoretical Physics,
                Faculty of Physics, St.Kliment Ohridski University of Sofia}\\
{\footnotesize  5, James Bourchier Blvd., 1164 Sofia, Bulgaria }\\
  {\footnotesize ${}^{4}$ Department of Physics, Aristotle University of Thessaloniki, Thessaloniki 54124, Greece}\\\\[-3.mm]
 Ivan Zh. Stefanov$^{5}$ \thanks{E-mail: izhivkov@yahoo.com}
 \\ [2.mm]{\footnotesize{${}^{5}$ Department of Applied Physics, Technical University of Sofia,}}\\ [-1.mm]{\footnotesize{8, Kliment Ohridski Blvd., 1000
Sofia, Bulgaria}}}

\date{}

\maketitle

\begin{abstract}
In the present paper, we study the scalar sector of the quasinormal modes of charged  general relativistic, static, and
spherically symmetric  black holes coupled to nonlinear electrodynamics and embedded  in  a class of scalar-tensor
theories. We find that for certain domain of the parametric space, there exist unstable quasinormal modes. The presence
of instabilities implies the existence of scalar-tensor black holes with primary hair that  bifurcate from the embedded
general relativistic black-hole solutions at critical values of the parameters corresponding to the static zero-modes. We prove that
such scalar-tensor black holes really exist by solving the full system of scalar-tensor  field equations for the
static, spherically symmetric case. The obtained solutions for the hairy black holes are non-unique, and they are in one
to one correspondence with the bounded states of the potential governing the linear perturbations of the scalar field.
The stability of the non-unique hairy black holes is also examined, and we find that the solutions for which the scalar field has zeros
are unstable against radial perturbations. The paper ends with a discussion of possible formulations of a new classification
conjecture.

\end{abstract}

%%%%%%%%%%%%%%%%%%%%%%%%%%%%%%%%%%%%%%%%%%%%%%%%%%%%%%%%%%%%%%%%%%%

%\draft
\sloppy

\section{Introduction}
The quasinormal  modes of black holes were intensively studied during the last decade. The great interest in the
quasinormal modes is motivated on the one side by   astrophysics and especially by the expected recent observation of
one of the most important predictions of general relativity -- the gravitational waves. Information about the physical
properties of the black holes (and other compact objects) can be obtained through  the emission of gravitational waves
and the characteristic frequencies of ringing, namely the frequencies of the quasi-normal modes (QNMs)
\cite{Kokkotas:1999bd, Nollert:1999ji}. The QNMs would allow us to distinguish between different compact objects --
black holes, stars, to distinguish between different theories of gravity since they predict different characteristic
spectra, to determine the global asymptotic charges like mass, charge, and angular momentum of the observed black holes
and so on \cite{Dreyer}.

On the other side, the quasinormal modes are interesting and important for intrinsic theoretical reasons.  During the
last  decade, the quasinormal modes  became a powerful tool for the study of various pure theoretical problems which deepen our
knowledge of spacetime and gravity, especially in the regime of strong fields \cite{cardosotopical}. They may serve, for example,
as an indicator of black-hole phase transitions \cite{RGCai1}--\cite{RGCai2}.

In the present paper, we study the scalar sector of the  quasinormal modes of  general relativistic, static, and
spherically symmetric black-hole solutions coupled to nonlinear electrodynamics which are also solutions to the field
equations of a certain class of scalar-tensor theories with nonlinear electrodynamics, i.e., they can be viewed as
embedded in the scalar-tensor theories under consideration. In this paper, we consider a class of scalar-tensor
theories defined by zero potential $V(\varphi)=0$ and a coupling function
\begin{eqnarray}
{\cal A}(\varphi)=e^{\frac{1}{2}\beta \varphi^2}
\end{eqnarray}
corresponding to $\alpha(\varphi)=\beta\varphi$  where $\beta$ is a constant\footnote{The general definition of the function
$\alpha(\varphi)$ is $\alpha(\varphi)=d\ln A(\varphi)/d \varphi$ and
more details can be found in the next section.}. These scalar-tensor theories\footnote{ In the present paper
we fix the cosmological value of the scalar field to be $\varphi_{\infty}=0$.} are indistinguishable from the general relativity
in weak field regime (the post-Newtonian parameters have the same values as in GR), but can differ
seriously from it in the strong field regime as it was first shown in \cite{DEFPRL} for neutron stars. It is natural to try to
find similar non-perturbative deviation from GR when black holes are considered as sources of strong gravitational filed. According to several
no-scalar-hair theorems, however, in the case
when no matter sources are present in the theory \cite{Saa} and in the case when the scalar-tensor theories are coupled to linear electrodynamics
\cite{BSen},
the static, spherically symmetric black-hole solutions are indistinguishable from the corresponding solutions in  GR
(see also \cite{Hawking} -- \cite{Heusler}). These theorems can be eluded when the scalar-tensor theories are coupled to non-linear electrodynamics
since it has a non-vanishing
trace of the energy-momentum tensor, and this nonzero trace acts as a source of the scalar field giving rise to black holes with nontrivial scalar field.
Such scalar-tensor black holes were first found in \cite{SYT1} and \cite{SYT1a}. Within the particular class of scalar-tensor
theories considered here, the static and spherically symmetric black holes coupled to nonlinear Born-Infeld electrodynamics were
studied in \cite{SYT2}. It was shown that black holes with nontrivial scalar field can exist only for $\beta<0$. The
most interesting result in  \cite{SYT2} is the fact that the considered class of scalar tensor theories with $\beta<0$
allows the existence of non-unique black-hole solutions with the same mass and charge. This disproves the no-hair conjecture
according to which the static scalar-tensor black holes are uniquely specified by their conserved
asymptotic charges, namely, the mass and the charge. In other words, the scalar-tensor black holes can support primary
scalar hair in contrast with the secondary scalar hair uniquely determined by the mass and the charge.

Among the non-unique solutions present in the considered theory is the solution with zero (trivial) scalar field which is just the
solution of the pure Einstein equations coupled to nonlinear Born-Infeld electrodynamics. From now on, the embedding of
this general relativistic solution  in the scalar-tensor theories under consideration will be called ``{\it the trivial solution}''. It turns
out that in a certain class of scalar-tensor theories, the perturbations of the trivial solution form two independent sectors -- scalar  and
metric-vector sector. In
other words, the perturbations of the scalar field decouple from the perturbations of the metric and electromagnetic
field. The equations governing the perturbations of the metric and the electromagnetic field coincide with those of
pure general relativity. Since we are interested  in the scalar-tensor theories, we focus only on the scalar sector.
The equation governing the scalar perturbations feels not only the background metric and background
electromagnetic field but also involves a term coming from the scalar-tensor theory under consideration. The presence
of this scalar-tensor term in the potential leads to interesting consequences.  Especially, the scalar-tensor term with
$\beta<0$ gives negative contribution to the potential of the scalar perturbations and a negative minimum in the
potential appear   for a certain domain of the parameter space. The bound states associated with the negative minimum of
the potential correspond to unstable modes. Therefore, for a certain sector of the parameter space, the trivial solution
becomes unstable. How the instability of the trivial solution should be interpreted? Our interpretation is  that the
instability is  a sign for the existence of new black-hole solutions with nontrivial primary scalar hair which
bifurcate from the trivial solution at critical parameters  corresponding to the static zero-modes. We prove this
interpretation  by numerically solving the full system of static, spherically symmetric scalar-tensor field equations and finding the
explicit solutions with nontrivial scalar field. The number of nontrivial  scalar-tensor black-hole solutions that bifurcate from the
trivial one turns out to be
in one to one correspondence with the number of bounded states of the scalar perturbations potential. In this way, we find a wide
spectrum of new solutions which is much richer than what was previously found in \cite{SYT2}.

We examine also  the stability of the nontrivial solutions. We show that the nontrivial solutions for which the scalar field has
zeros are unstable against radial perturbations.

The paper is organized as follows. In section 2, we introduce the necessary background of general equations. In section
3,
the equation governing the scalar perturbations is derived. The numerical methods for the calculations of the QNMs and the numerical results are
presented in section 4. The spherically symmetric solutions describing hairy black holes and bifurcating from the trivial solution
are found numerically in section 5. In section 6,  we examine the stability of the nontrivial black-hole solutions. The paper ends with
 discussion.

\section{General equations}

The action of the  scalar-tensor theories coupled to nonlinear electrodynamics in the Einstein frame is
\begin{eqnarray}
&&S=\frac{1}{16\pi G_*} \int{d^4 x \sqrt{-g}\left(R - 2g^{\mu\nu}\partial_\mu \varphi \partial_\nu \varphi - 4V(\varphi)\right)} \notag \\ \notag \\
&&\hspace{0.9cm} + \frac{1}{4\pi G_*} \int{d^4 x \sqrt{-g} {\cal A}^4(\varphi)L(X,Y) }, \label{eq:Action_EF}
\end{eqnarray}
where $G_*$ is the bare gravitational constant, $g$ is the determinant of the Einstein frame metric $g_{\mu\nu}$, $R$
is the Ricci scalar curvature with respect to the metric $g_{\mu\nu}$, $\varphi$ is the scalar field, $V(\varphi)$ is
the potential of the scalar field, ${\cal A}(\varphi)$ is the coupling function between the sources of gravity and the
scalar field and $L(X,Y)$ is the Lagrangian of the nonlinear electrodynamics. The functions $X$ and $Y$ are given by
\begin{eqnarray}
&&X=\frac{{\cal A}^{-4}(\varphi)}{4}F_{\mu\nu}F^{\mu\nu}, \label{f:eqX}\\ \notag \\
&&Y=\frac{{\cal A}^{-4}(\varphi)}{4}F_{\mu\nu}(\star F)^{\mu\nu},
\end{eqnarray}
where $F_{\mu\nu}$ is the electromagnetic field strength tensor and ``$\star$'' stands for the Hodge dual with respect
to the metric $g_{\mu\nu}$.

The field equations derived from the action are given by
\begin{eqnarray}
&&R_{\mu\nu} = 2\partial_{\mu}\varphi \partial_{\nu}\varphi +  2V(\varphi)g_{\mu\nu} -
 2\partial_{X} L(X, Y) \left(F_{\mu\beta}F_{\nu}^{\beta} -
{1\over 2}g_{\mu\nu}F_{\alpha\beta}F^{\alpha\beta} \right)  \nonumber \\
&&\hspace{2cm}-2{\cal A}^{4}(\varphi)\left[L(X,Y) -  Y\partial_{Y}L(X, Y) \right] g_{\mu\nu}, \nonumber  \\ \nonumber \\
&&\nabla_{\mu} \left[\partial_{X}L(X, Y) F^{\mu\nu} + \partial_{Y}L(X, Y) (\star F)^{\mu\nu} \right] = 0 \label{eq:FieldEq},\\ \nonumber \\
&&\nabla_{\mu}\nabla^{\mu} \varphi = {d V(\varphi)\over d\varphi } -
4\alpha(\varphi){\cal A}^{4}(\varphi) \left[L(X,Y) -  X\partial_{X}L(X,Y) -  Y\partial_{Y}L(X, Y) \right], \nonumber
\end{eqnarray}
where
\begin{equation}
\alpha(\varphi) = {d \ln{\cal A}(\varphi)\over d\varphi}. \label{eq:alpha_def}
\end{equation}
In the present paper, we consider a class of scalar-tensor theories defined by zero potential $V(\varphi)=0$ and a
coupling function
\begin{eqnarray}
{\cal A}(\varphi)=e^{\frac{1}{2}\beta \varphi^2}
\end{eqnarray}
corresponding to $\alpha(\varphi)=\beta\varphi$ where $\beta$ is a constant. According to some estimations the
parameter $\beta$ must obey the inequality $\beta>-4.5$ in order for the theory to be consistent with the experimental
data \cite{DFBPE}. That is why we shall restrict our considerations in this region of values of $\beta$. The scalar-tensor theories
with $\alpha(\varphi)=\beta\varphi$ (and cosmological value $\varphi_{\infty}=0$) as we have already mentioned are indistinguishable
from general relativity in the weak field regime but can exhibit non-perturbative effects
and can differ seriously from general relativity in the strong field regime \cite{DEFPRL},\cite{SYT2}.

In order to  be specific, we will consider here the  Born-Infeld nonlinear electrodynamics. An important property of
this nonlinear electrodynamics is that it is invariant under electric-magnetic duality rotation so it is enough to
consider only magnetically charged solutions, and the electrically charged can be simply obtained from them. For pure
magnetic solutions, $Y=0$ and the truncated Born-Infeld Lagrangian is
\begin{equation}
L(X) = 2b \left( 1- \sqrt{1+ \frac{X}{b}} \right)\label{eq:BILagran}
\end{equation}
where $b$ is a parameter and in the limit $b\rightarrow \infty$ the Maxwell electrodynamics is restored. It can be easily shown that for
the Born-Infeld theory, the following inequality holds
\begin{equation}
X \partial_{X}L(X) - L(X)>0. \label{eq:BIInequality}
\end{equation}

We will consider static, spherically symmetric,  asymptotically flat black holes and the ansatz for the metric is
\begin{equation}
ds^2 = g_{\mu\nu}dx^{\mu}dx^{\nu} = - f(r)e^{-2\delta(r)}dt^2 + {dr^2\over f(r) } +
r^2\left(d\theta^2 + \sin^2\theta d\phi^2 \right). \label{eq:metric}
\end{equation}
In this case, the function $X$  reduces to
\begin{equation}
X = {{\cal A}^{-4}(\varphi)\over 2} \frac{P^2}{r^4},
\end{equation}
where $P$ is the magnetic charge \cite{SYT2,SYT1}. Respectively, the field equations are reduced to the following
system of ordinary differential equations \cite{SYT2,SYT1}
\begin{eqnarray}
&&\frac{d\delta}{dr}=-r\left(\frac{d\varphi}{dr} \right)^2 \label{eq:ODEDelta},\\
&&\frac{d m}{dr}=r^2\left[\frac{1}{2}f\left(\frac{d\varphi}{dr} \right)^2 -
{\cal A (\varphi)}^{4}L(X)  \right] \label{eq:ODEMass},\\
&&\frac{d }{dr}\left( r^{2}f\frac{d\varphi }{dr} \right)=
r^{2}\left\{-4\alpha(\varphi){\cal A}^{4}(\varphi) \left[L -  X\partial_{X}L(X)\right] -
r f\left(\frac{d\varphi}{dr} \right)^3    \right\} \label{eq:ODEPhi}  ,
\end{eqnarray}
where we have introduced the new function $m(r)$ defined by $f(r)=1-2m(r)/r$. Without loss of generality we consider
scalar fields that vanish at infinity. Taking into account that $\alpha(\varphi)=\beta \varphi$ it is not difficult to
see that every static, spherically symmetric solution of the Einstein equations coupled to a nonlinear electrodynamics is
also a solution to the above system with  $\varphi(r)\equiv0$. The general relativistic solution describing black hole
coupled to a nonlinear electrodynamics is naturally embedded in the class of scalar-tensor theories under consideration
and, as we have already mentioned, will be called {\it the trivial solution}. The metric functions of the trivial
solution are given by

\begin{eqnarray}
&&\delta(r)=0 \notag\\
&&f(r)=1 - \frac{2M}{r} + \frac{4 b r^2}{3} \left(1-\sqrt{1+\frac{P^2}{2 b r^4}}\right)  \label{eq:f(r)_EBI}\\
&& \hspace{1.3cm} + \frac{4P^2}{3r^2} {\cal F}\left(\frac{1}{4},\frac{1}{2},\frac{5}{4},-\frac{P^2}{2 b r^4}\right)  \notag
\end{eqnarray}
where ${\cal F}$ is the hypergeometric function \cite{Abramowitz} and $M$ is the black hole mass.

For the trivial solution, it can be easily proven  that extremal black-hole solutions can exist only for $bP^2>1/8$.
However, no extremal black holes can exists in the considered class of scalar-tensor theories when the scalar field is
nontrivial\footnote{A detailed proof can be found in \cite{SYT2}}.

\section{Perturbations of the trivial solution}

In the present paper, we consider the perturbations of the trivial solution within the framework of the described  class
of scalar-tensor theories. The metric perturbations will be  denoted by
$\delta g_{\mu\nu}$, the electromagnetic perturbations\footnote{Here, $A_{\mu}$ is the gauge potential, i.e.,
$F_{\mu\nu}=\partial_{\mu}A_{\nu}- \partial_{\nu}A_{\mu}$.} by $\delta A_{\mu}$ and the scalar ones by $\delta\varphi$.
Since the trivial solution has zero scalar field, it is not difficult to see that in the considered class of
scalar-tensor theories the  equations governing the perturbations of the metric $\delta g_{\mu\nu}$ and the
electromagnetic field $\delta A_{\mu}$ are decoupled from the equation governing the perturbations $\delta\varphi$ of
the scalar field. The equations for  $\delta g_{\mu\nu}$ and $\delta A_{\mu}$   are in fact the same as those in the
pure Einstein gravity. Since we are interested in black holes in the scalar-tensor theories we will focus only on
the scalar perturbations $\delta\varphi$.

From the third  field equation in  (\ref{eq:FieldEq}), we obtain the following equation governing the
perturbations of the scalar field
\begin{eqnarray}\label{pereq1}
\nabla^{(0)}_\mu \nabla^{(0)\mu} \delta \varphi=4\beta  [X^{(0)} \partial_{X^{(0)}}L\left(X^{(0)}\right) - L\left(X^{(0)}\right)] \delta \varphi, \label{eq:PertFieldEq}
\end{eqnarray}
where the covariant derivative $\nabla^{(0)}_\mu$ and the function $X^{(0)}$ are with respect to the unperturbed background solution. Since the
background solution is static and spherically symmetric, the variables can be separated in the following way
\begin{align}
\delta \varphi ={ \psi(r)\over r} e^{i\omega t}Y_{lm}(\theta,\phi), \label{eq:SeparVar}
\end{align}
where $\omega$ is a parameter and $Y_{lm}(\theta,\phi)$ are the spherical harmonics. After substituting in
(\ref{pereq1}) we find
\begin{eqnarray}\label{pereq2}
f\frac{d}{dr}\left(f\frac{d\psi}{dr}\right)+[\omega^2-U(r)]\psi=0,\label{eq:PertEqOrig}
\end{eqnarray}
where the potential $U$ is defined by
\begin{align}
U(r)=f\left[\frac{1}{r}\frac{df}{dr}+\frac{l(l+1)}{r^2} + 4\beta[X^{(0)} \partial_{X^{(0)}}L\left(X^{(0)}\right) - L\left(X^{(0)}\right)] \right]. \label{eq:potentU}
\end{align}
The equation (\ref{pereq2}) can be cast in the Schr\"odinger  form
\begin{eqnarray}
\frac{d^2 \psi(r_*)}{dr^2_*}+\omega^2 \psi(r_*)=U(r_*)\psi(r_*) \label{eq:SchrodTypeEq}
\end{eqnarray}
by introducing the tortoise coordinate
\begin{align}
dr_*=\frac{dr}{f(r)} \label{eq:change2}
\end{align}
which maps the domain $r \in [r_H,\infty)$ ($r_H$ is the radius of the horizon of the black hole) to $r_* \in
(-\infty,\infty)$.

\section{Quasi-normal modes}

\subsection{Dimensionless quantities}

The reduced system of equations (\ref{eq:ODEDelta})--(\ref{eq:ODEPhi}) is invariant under the rigid rescaling
\begin{equation}
r\rightarrow \lambda r,  \,\, \,\, m\rightarrow \lambda m,  \,\, \,\, b\rightarrow \lambda^{-2} b,  \,\, \,\, P\rightarrow \lambda P,
\end{equation}
where  $\lambda \in (0,\infty)$. Therefore, one may generate in this way a family of new solutions. The mass, the
temperature, and the entropy of the new solutions are given by the formulas
\begin{eqnarray}
M\rightarrow \lambda M,  \,\,  \,\, T\rightarrow \lambda^{-1} T,  \,\, \,\, S\rightarrow \lambda^2 S .
\end{eqnarray}

Respectively, the equation for the perturbations of the scalar field (\ref{eq:PertEqOrig}) is invariant under the rigid rescaling
\begin{equation}
r\rightarrow \lambda r, \,\, \,\, m\rightarrow \lambda m, \,\, \,\, b\rightarrow \lambda^{-2} b, \,\, \,\,  P\rightarrow \lambda P,\,\, \,\, \omega \rightarrow \lambda^{-1} \omega .
\end{equation}

As a consequence of the rigid symmetry we may restrict our study to the case $b=0.01$ by introducing the dimensionless
quantities

\begin{equation}
r\rightarrow \frac{r}{\xi},  \,\, \,\, m\rightarrow \frac{m}{\xi},  \,\, \,\, b\rightarrow b\xi^2,  \,\, \,\, P\rightarrow \frac{P}{\xi}, \,\, \,\,
\omega \rightarrow \xi \omega, \,\,\,\, M\rightarrow \frac{M}{\xi},
\end{equation}
where the choice\footnote{The very choice of the value 0.01 is for numerical convenience.} $b=0.01$ requires
\begin{equation}
\xi=\frac{0.1}{\sqrt{b}}.
\end{equation}

\subsection{Qualitative considerations and numerical algorithm}
Our task is to calculate the QNMs of the trivial solution embedded  in the considered class of scalar-tensor theories.
Thus we have to find the QNM frequencies governed by eq.
(\ref{eq:PertEqOrig}) (or equivalently by eq. (\ref{eq:SchrodTypeEq})) where the boundary condition are the standard
ones: purely outgoing wave at infinity and purely ingoing wave at the black
hole horizon. We have to note that the problem reduces to studying the scalar perturbations of the pure Einstein-Born-Infeld black holes when the
parameter in the coupling function $\beta=0$ and this case is studied in \cite{Fernando} \footnote{Our results for $\beta=0$ are
qualitatively similar to the results obtained by Fernando and Holbrook in \cite{Fernando} but the computed numerical values of the frequencies are different.}.

\begin{figure}[t]%
\vbox{ \hfil \scalebox{1.00}{ {\includegraphics{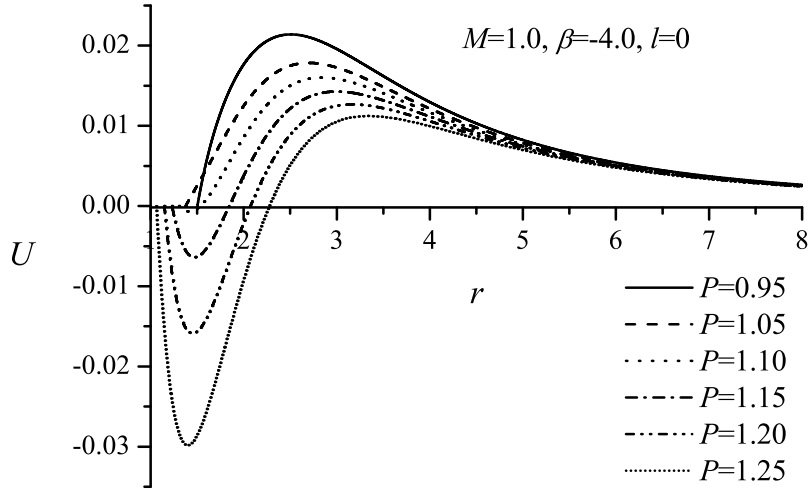}} }\hfil}%
\caption{The potential $U$ as a function of $r$ for values of the parameters $M=1.0$, $\beta=-4.0$, $l=0$ and for several values of the magnetic
charge $P$. The range of values of $P$ is chosen so that the formation of a negative minimum can be seen as we increase the charge.} \label{Fig:U(r)}
\end{figure}

It is the potential $U$ that determines the  qualitative and quantitative behavior of the  QNMs for the already fixed boundary conditions.
That is why  we will first comment on the qualitative behavior of $U$ defined by eq. (\ref{eq:potentU}). It can be easily checked that the only difference between the
equation governing the perturbations of the scalar field (\ref{eq:PertEqOrig}) and the equation governing the scalar perturbations in the case of
pure Einstein-Born-Infeld theory is the term $4\beta[X^{(0)}
\partial_{X^{(0)}}L\left(X^{(0)}\right) - L\left(X^{(0)}\right)]$ in the potential (\ref{eq:potentU}). When we take into account equation
(\ref{eq:BIInequality}), it is clear that the sign of this term is the same as the sign of $\beta$. Thus, the potential
can become negative when $\beta<0$. On Fig. \ref{Fig:U(r)}, the potential $U(r)$ for $M=1.0$, $\beta=-4$ and $l=0$  is
plotted for a range of parameters. As it can be seen, a positive maximum is present for all curves, and the numerical
results show that such maximum also exists for all of the studied values of the parameters $M$, $P$, $\beta$, and $l$.
For large values of the charge, an additional negative minimum appears when $\beta<0$,  and it can lead to instabilities.
When $\beta>0$, the potential is qualitatively the same as for the pure Einstein case.

For fixed values of the parameters, the depth of the potential well, if it exists, decreases with the increase of $l$ because of the
contribution of the second term  in (\ref{eq:potentU}). For high enough
$l$, the potential becomes positive.

The calculation of the QNMs in practice, however, faces serious numerical difficulties. The first reason for that is the complexity
of the background solution which can not be expressed in elementary functions as well as the presence of negative minimum in the potential
for some values of the parameters. This  makes us  use  direct integration methods which has limited abilities since the  more sophisticated and accurate methods are unapplicable.  The second reason is the numerical instabilities that appear in the region of transition from stable to unstable modes which will be discussed below. It is well known that even for the Schwarzschild solution, the
direct integration methods face serious difficulties when the imaginary part of the quasinormal frequencies becomes greater than the real part.

It turns out that the properties of the perturbation functions and the modes in the stable and unstable
sector are different, and that is why it is useful to consider the stable and the unstable modes separately. More precisely,
as we shall see below, the unstable modes define a self-adjoined Sturm-Liouville boundary value problem for which the
eigenvalues are real. This simplifies the treatment of the problem and makes the numerical procedure  stable.

\subsubsection{Stable modes} \label{sec:StableQNM}
The required QNMs boundary conditions lead to an exponential growth of the radial perturbation function for the stable modes as $r_*\rightarrow \pm\infty$ .
To circumvent this difficulty, we use the numerical procedure introduced by Chandrasekhar and Detweiler
\cite{ChandraShooting},\cite{KokkotasRN}. The idea of the method is to introduce a new function $\phi(r_*)$ defined by
\begin{equation}
\psi = e^{i \int{\phi dr_*}}, \label{eq:DefinePhase}
\end{equation}
where $\psi$ is the radial part of the perturbations of the scalar field. Equation (\ref{eq:SchrodTypeEq})
transforms into a Riccati type equation
\begin{equation}
i\frac{d\phi}{dr_*} + \omega^2 - \phi^2 - U(r_*) = 0 \label{eq:PertEqShooting}
\end{equation}
and the boundary conditions, according to the sign convention in (\ref{eq:SeparVar}), are
\begin{equation}
\phi|_{r_* \rightarrow -\infty}\rightarrow \omega ,\hspace{1cm} \phi|_{r_* \rightarrow \infty}\rightarrow -\omega. \label{eq:BC_shooting}
\end{equation}

We apply the following shooting procedure for calculating the QNM frequencies. Equation (\ref{eq:PertEqShooting}) is
integrated for test values of the  complex parameter $\omega$ and a QNM frequency $\omega$ is found when the boundary conditions
(\ref{eq:BC_shooting}) are satisfied within the required accuracy. For each test value of $\omega$ two integrations
are performed -- inward and outward, and they are described below in details.

First, we integrate equation (\ref{eq:PertEqShooting}) outward from a chosen point $r_{*-\infty}$ to a matching point
$r_{* \rm{match}}$ which is close to the maximum of the potential. The initial condition at $r_{* -\infty}$ is computed using
the expansion of the function $\psi$ at the black hole horizon
\begin{equation}
\psi|_{r_* \rightarrow -\infty} = e^{i\omega r_*}\sum^{\infty}_{s=1} {\beta_s (r-r_H)^{s-1}}, \label{eq:BC_rh}
\end{equation}
where $r_H$ is the radius of the horizon and $\beta_s$ are constants ($\beta_1=1$ and $\beta_s=0$ for $s<1$). The
coefficients $\beta_s$ ($s>1$) can be calculated from the recurrence relation that is obtained when we substitute (\ref{eq:BC_rh}) into
eq. (\ref{eq:SchrodTypeEq}) .

The second inward integration is from a chosen point $r_{* \infty}$ \footnote{The point $r_{* \infty}$
is chosen so that the ratio $r_{* \infty}/|\omega|$ is kept fixed.} to the same matching point $r_{*
\rm{match}}$ and the initial condition at $r_{* \infty}$ is calculated from the asymptotic expansion at infinity
\begin{equation}
\psi|_{r_* \rightarrow \infty} = e^{-i\omega r_*}\sum^{\infty}_{s=1} {\alpha_s r^{1-s}} \label{eq:BC_inf}
\end{equation}
where $\alpha_s$ are constants ($\alpha_1=1$ and $\alpha_s=0$ for $s<1$). The coefficients $\alpha_s$ ($s>1$) can be
found from a recurrence relation derived in the same way as for the $\beta_s$ .

After the two integrations, we compare the values of the function $\phi$, resulting from the inward and outward
integration, at the matching point $r_{* \rm{match}}$, and a QNM frequency $\omega$ is found when these values coincide
within the required accuracy.

As it is pointed out in \cite{ChandraShooting}, the numerical method we use may become unstable when the imaginary part of the QNM frequency becomes
greater that the real one, and this situation actually occurs in our calculations in the region of transition from stable to unstable modes.
That is why we have to use a stiff ODE solver, and during our tests, it turned
out that the semi-implicit extrapolation method described in Numerical Recipes \cite{NR} is a very good choice. The search of  QNM
frequencies $\omega$ described above, can be considered as a root search procedure,  and we used the Muller method for it \cite{NR}.

\subsubsection{Unstable modes} \label{sec:UnstableQNM}
According to our sign convention in (\ref{eq:SeparVar}), the unstable QNM frequencies
should have negative imaginary part. In this case, we can substitute $\omega=\omega_{R} + i \omega_I$, where
$\omega_{R}$ and $\omega_I$ are both real and $\omega_I$ is negative. Hence, the boundary conditions for the unstable
quasi-normal modes are vanishing on both ends
\begin{eqnarray}
&&\psi|_{r_* \rightarrow -\infty}\rightarrow e^{i\omega_R r_*} e^{-\omega_I r_*}|_{r_* \rightarrow -\infty}\rightarrow 0, \\
&&\psi|_{r_* \rightarrow \infty}\rightarrow e^{-i\omega_R r_*} e^{\omega_I r_*}|_{r_* \rightarrow \infty}\rightarrow 0.
\end{eqnarray}
But in this case the boundary value problem (BVP) discussed above is self-adjoint and has real eigenvalues $\omega^2$. Therefore, the unstable
QNM frequencies should be purely imaginary, and they correspond to the bound states of the potential well. More information on this
subject can be found in \cite{TornBook}, and similar considerations, but for neutron starts in scalar-tensor theories, are made in \cite{Harada}.

The perturbation function $\psi$ for the unstable modes will have no zeros for the first bound state,  one
zero for the second bound state, and so on.
The function $\phi$ defined by (\ref{eq:DefinePhase}) is divergent in the points where $\psi$ is equal to zero which makes
the numerical procedure used by Chandrasekhar and Detweiler inapplicable  for calculating the unstable modes (except
for the first bound state where $\psi$ has no zeros). Hence, we have to solve the original equation for the $\psi$
function (\ref{eq:SchrodTypeEq}) and the same expansions on the boundaries (\ref{eq:BC_rh}) and (\ref{eq:BC_inf}) can be
used for calculating the boundary conditions.

The numerical method used is quite similar to the shooting procedure described above: equation (\ref{eq:SchrodTypeEq}) is integrated inward and
outward for test values of the parameter $\omega$, and because of the arbitrary linear scale, it is enough to match
at the chosen point $r_{* \rm{match}}$ only the quantities
\begin{equation}
\frac{1}{\psi} \frac{d\psi}{dr_*}
\end{equation}
resulting from the two integrations.

\subsection{Results}
We calculate the QNM frequencies of the trivial solutions using the numerical techniques described above. As we said, we will restrict
ourselves to calculating the QNM frequencies for $\beta>-4.5$ which is required from the observational constraints. The  presented  results
are for non-extremal black holes, i.e., for $bP^2<1/8$. Thus, the end
of the sequences of black-hole solutions is reached when $r_H\rightarrow 0$ which, for example, for $M=1$ corresponds to charge
$P\approx 1.67$. The results in the case when  extremal black holes are present(i.e., for $bP^2>1/8$) are qualitatively the same.

Before proceeding further, we will give some remarks on the computational abilities of the numerical method we use. As the numerical
method is unstable in the region where the imaginary part of $\omega$ is greater that the real part, we can compute only the
fundamental ($n=0$) QNM frequency when $l=0$, because even for Schwarzschild black hole,
the $n=1$, $l=0$ mode is $\omega_{n=1} = 0.086 + 0.348 i$ and the ratio of the imaginary part to the real part is
beyond the computational abilities of the method we use.

\begin{figure}[t]%
\includegraphics[width=0.50\textwidth]{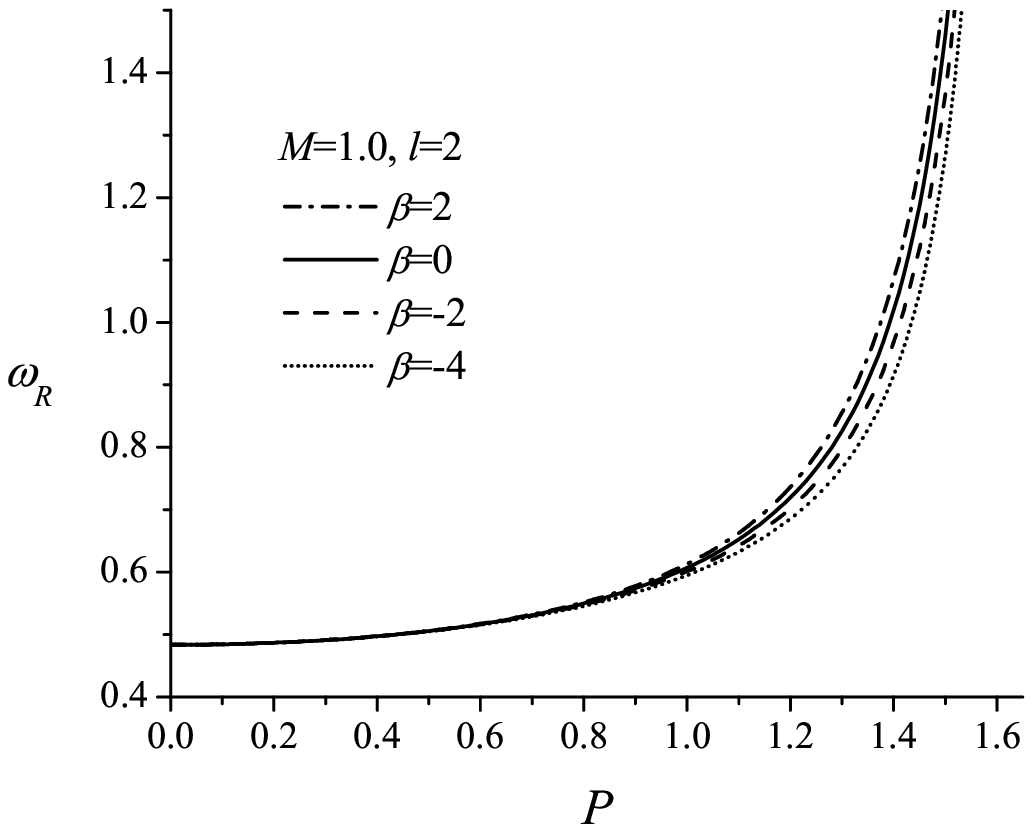}
\includegraphics[width=0.53\textwidth]{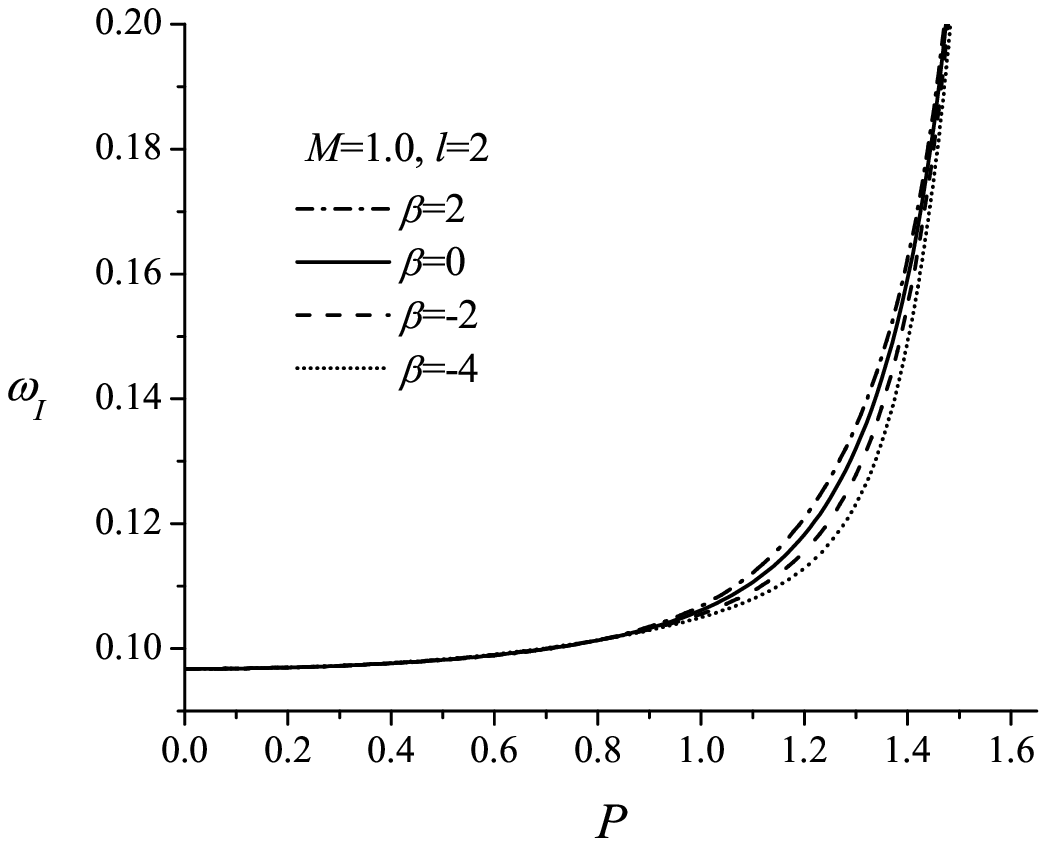}
\caption{The real $\omega_R$ (left panel) and the imaginary $\omega_I$ (right panel) part of the $l=2$ fundamental ($n=0$) frequencies
as a function of the charge for several values of the parameter $\beta$ and for $M=1.0$.}
\label{fig:ReIm(P)l2M1}
\end{figure}

The results for the  $l=2$ fundamental modes as functions of the charge are shown on Fig. \ref{fig:ReIm(P)l2M1} for several
values of the parameter $\beta$ and for $M=1$, and as we can see, both the real and the imaginary part of the modes increase as we
increase the charge~\footnote{For $M=1$ and $\beta=-4$, the potential is positive for $l\geq 1$.}.
The results are qualitatively the same for other values of
$l\ge 1$. The positivity of the imaginary part
according to our sign convention (\ref{eq:SeparVar}) means that the modes are damped and this is expected from the fact that the potential is positive
for this values of the parameters.

\begin{figure}[t]%
\includegraphics[width=0.50\textwidth]{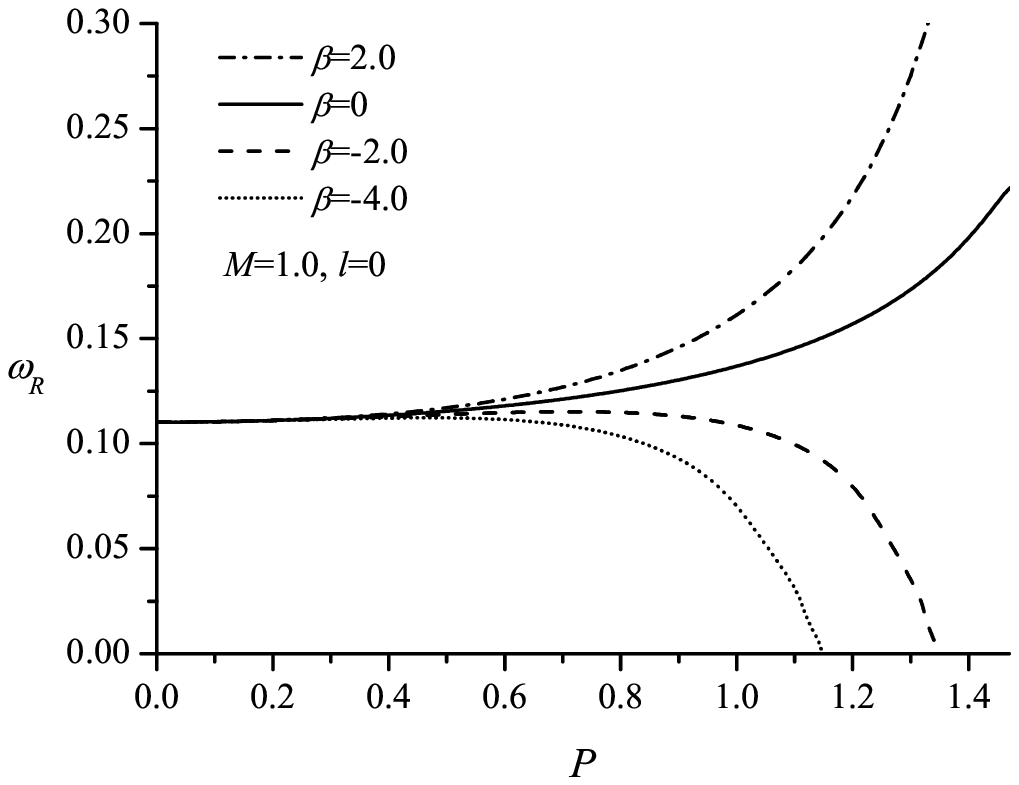}
\includegraphics[width=0.53\textwidth]{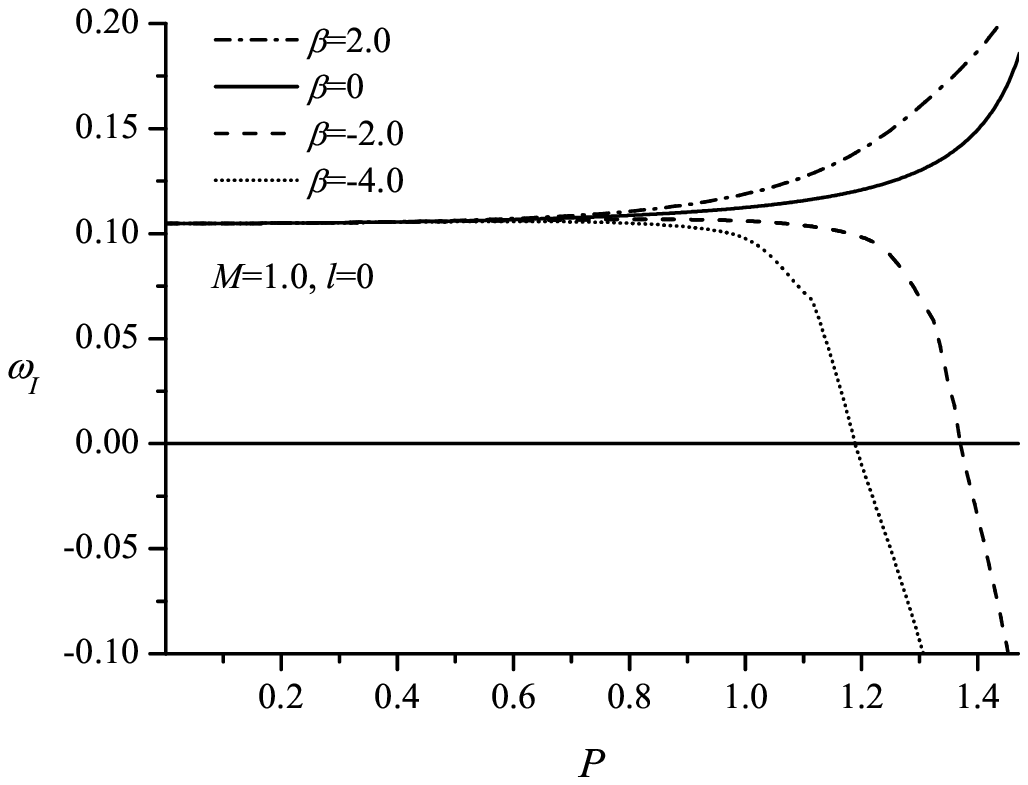}
\caption{%
The real $\omega_R$ (left panel) and the imaginary $\omega_I$ (right panel) part of the $l=0$ fundamental QNM frequencies as a function of the
magnetic charge $P$ for several values of the parameter $\beta$ and for $M=1$.} \label{fig:Omega_VarBeta}%
\end{figure}

The situation changes when $l=0$ because in this case a negative minimum of the potential can exist when $\beta<0$. On Fig.
\ref{fig:Omega_VarBeta}, the fundamental $l=0$ QNM frequencies are shown as a function of the charge for $M=1$ and for
several values of the parameter $\beta$. Some of the calculated numerical values for the QNM frequencies are presented in Table \ref{tbl:TableQNMs}.
The real and the imaginary part of the frequencies increase with the increase of the
charge for $\beta\geq0$. But
for $\beta<0$ and for large values of the charge, both the real and the imaginary part of the frequencies start to
decrease and eventually reach zero at some critical charge $P^{(0)}_{\rm crit}$ \footnote{The critical charges where the real and the
imaginary part reach zero coincide within $3\%$ and the small discrepancy is due to the
numerical instabilities described below.}. For larger values of the charge the real part is zero and the imaginary
part is negative which means that the perturbations of the scalar field are growing with time and the corresponding black-hole solution is unstable.

\begin{table*}[t!]
\begin{center}
\caption{The calculated numerical values of the $l=0$ fundamental QNM frequencies. Results for $M=1$ and for several values of the
parameter $\beta$ and the charge $P$ are given.} \label{tbl:TableQNMs}
\begin{tabular}{|c|cc|c|cc|c|cc|c|cc|}
\hline
\multicolumn{1}{|c|}{ } & \multicolumn{2}{|c|}{$\beta=2$} &  & \multicolumn{2}{|c|}{$\beta=0$} &  & \multicolumn{2}{|c|}{$\beta=-2$} &  & \multicolumn{2}{|c|}{$\beta=-4$}\\
\hline
\hline
$P$ & $\omega_R$ & $\omega_I$ &  & $\omega_R$ & $\omega_I$ &  & $\omega_R$ & $\omega_I$ &  & $\omega_R$ & $\omega_I$\\
\hline
0.00 & 0.1102 & 0.1049 &  & 0.1102 & 0.1049 &  & 0.1102 & 0.1049 &  & 0.1102 & 0.1049  \\
\hline
0.20 & 0.1111 & 0.1050 &  & 0.1111 & 0.1050 &  & 0.1110 & 0.1050 &  & 0.1110 & 0.1050 \\
\hline
0.40 & 0.1142 & 0.1056 &  & 0.1135 & 0.1056 &  & 0.1129 & 0.1055 &  & 0.1122 & 0.1054 \\
\hline
0.60 & 0.1211 & 0.1071 &  & 0.1180 & 0.1067 &  & 0.1148 & 0.1062 &  & 0.1114 & 0.1058 \\
\hline
0.80 & 0.1349 & 0.1106 &  & 0.1252 & 0.1087 &  & 0.1149 & 0.1068 &  & 0.1035 & 0.1049 \\
\hline
1.00 & 0.1613 & 0.1189 &  & 0.1369 & 0.1124 &  & 0.1087 & 0.1060 &  & 0.0703 & 0.0976 \\
\hline
1.20 & 0.2177 & 0.1401 &  & 0.1570 & 0.1208 &  & 0.0796 & 0.0984 &  & 0 & -0.0100 \\
\hline
1.40 & 0.3788 & 0.1868 &  & 0.1980 & 0.1494 &  & 0 & -0.0346 &  & 0 & -0.2097 \\
\hline
1.55 & 0.7006 & 0.3122 &  & 0.2030 & 0.2160 &  & 0 & -0.3161 &  & 0 & -0.6487 \\
\hline
\end{tabular}
\end{center}
\end{table*}

In the region where the real and the imaginary part of the frequency decreases, the imaginary part of the frequency is
greater that the real one and the difference between them gets greater as we approach the critical charge $P^{(0)}_{\rm crit}$ which
means that in this region numerical instabilities are present.  Indeed, the numerical results show that the computed values of
the frequencies in this region depend on the values of the auxiliary parameters, such as $r_{*\infty}$ and $r_{* \rm{match}}$, but the difference
is up to $5\%$.
Another justification of our results is the following. The critical value of the magnetic charge $P^{(0)}_{\rm crit}$ can be
calculated in two independent ways using the imaginary part of the modes in the stable and the unstable sectors because the method
of calculation of the modes in each of them is different.
Much to our pleasant surprise, the results for $P^{(0)}_{\rm crit}$ obtained with the two independent methods coincide within $1\%$.

\begin{figure}[t]%
\vbox{ \hfil \scalebox{0.80}{ {\includegraphics{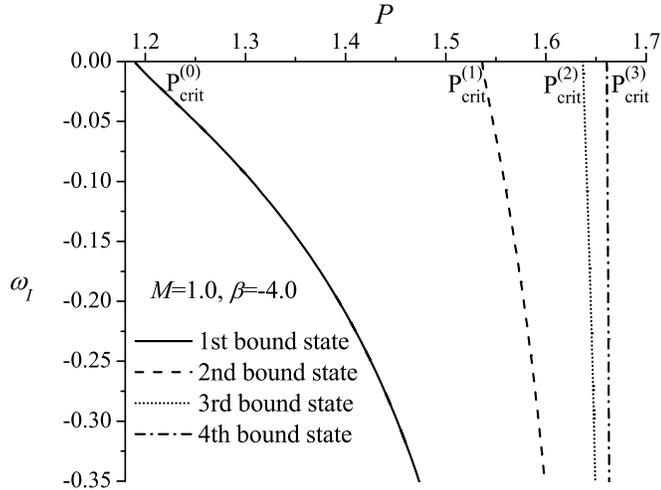}} }\hfil}%
\caption{The imaginary part $\omega_I$ as a function of the charge for the first four bound states ($\omega_R=0$ in this case).} \label{fig:BoundStateM1}
\end{figure}

The unstable QNM frequencies which are shown on Fig. \ref{fig:Omega_VarBeta}, correspond only to the first bound state of the
potential well. With the increase of $P$, the potential well gets deeper, and the number of bound states (unstable modes) increases.
We will label the different bound states with $k$, where $k=0,1,2...$ . The $k$-th unstable mode originates at $P_{\rm crit}^{(k)}$,
where $P_{\rm crit}^{(0)}<P_{\rm crit}^{(1)}<P_{\rm crit}^{(2)}<...$, and the radial part of the corresponding perturbation function
has $k$ zeros. The branches of the unstable modes of the first few bound states, are shown on Fig.~\ref{fig:BoundStateM1} for $M=1$ and $\beta=-4$.

It is clear that the first bound state (with $k=0$) is a continuation of the $n=0$ QNM
frequencies into the unstable region on the $\omega_I(P)$ diagram. We expect that the second bound state (with $k=1$) is a continuation
of the $n=1$ QNM frequencies and so on, but we can not check this numerically because as we said for $l=0$ only the fundamental $n=0$ mode
can be calculated by the applied numerical method.

\section{Bifurcations and black hole non-uniqueness}
The presence of unstable modes (for $l=0$) gives us a reason to expect that new spherically symmetric black-hole solutions
with nontrivial scalar field exist and should bifurcate from the trivial solution. Some examples of such  black-hole solutions have already been
found in \cite{SYT2}. In the phase diagrams in \cite{SYT2}, however, only one bifurcation point is present, and our numerical results show that it is just the point where the $n=0$ mode reaches zero \footnote{In \cite{SYT2}, the charge $P$ is kept fixed
and the mass $M$ is varied. With the decrease of $M$, the properties of the solutions change in the same way as in the case considered here when $M$ is kept fixed and $P$ is increased.}. The scalar field of
the non-trivial solution in \cite{SYT2} that bifurcates from the trivial solution is monotonous function of the radial coordinate, and when we take into
account that the scalar field is zero at infinity, it is clear that it  has
no zeros.  In the more general context of the present paper, it is natural to expect that solution bifurcations occur
for all of the critical charges $P_{\rm crit}^{(k)}$ where branches of unstable modes originate. In other words, bifurcations
should occur for the same values of the charge $P$ for which the frequencies of the unstable modes become zero. The
number\footnote{We mean the number of nontrivial solutions up to the obvious discrete $Z_2$ symmetry
$\varphi\rightarrow -\varphi$ of the field equations.} of the nontrivial solutions
that bifurcate from the trivial solution  should be equal
to the number of the unstable modes -- as we increase $P$, the number of bound states increases, and for each bound state that is added to the spectrum, a new bifurcation point occurs on the phase diagram at the same value of $P$.
Actually,  the static  zero-modes (i.e., modes with  $\omega=0$ and $l=0$) are perturbative solutions of the field equations describing static,
spherically symmetric black holes with non-trivial scalar field (i.e. hairy black holes) in the near vicinity of the bifurcation points.
Consequently, we should expect that the scalar field of the non-trivial black-hole solutions originating at the bifurcation points should
have the same number of zeros as a function of $r$ as the corresponding unstable zero-mode.

In order to check our hypothesis, we should  solve the full system of field equations  (\ref{eq:ODEDelta}) -- (\ref{eq:ODEPhi}).
For this we will follow a modified form of the  procedure from \cite{SYT2}. The system is
coupled and nonlinear, and it has to be solved numerically. The domain of integration is from the radius of the horizon
to infinity $r \in [r_H,\infty)$ and the boundary conditions on both boundaries are
\begin{equation}
f(r_H)=0,\label{cond_horizon}
\end{equation}
\begin{eqnarray}
&&\lim_{r \to \infty}m(r) =M, \label{eq:BC_delta} \\
&&\lim_{r \to \infty}\delta(r)=\lim_{r \to \infty}\varphi(r)=0\label{eq:BC_phi_delta},
\end{eqnarray}
where $M$ is the mass of the black hole in the Einstein frame. The following regularization condition should also be
fulfilled on the horizon
\begin{equation}
\left\{\left.\frac{df}{dr}\!\cdot\! \frac{d \varphi}{d r}- 4 \alpha(\varphi) {\cal A}^4(\varphi) [X \partial_X L(X)-L(X) ]\right\}\right|_{r=r_H}=0.
\label{regul}
\end{equation}

The system (\ref{eq:ODEDelta}) -- (\ref{eq:ODEPhi}) together with the boundary conditions is a BVP.
The mass is an input parameter, and the radius of the horizon has to be determined additionally which means that
the left boundary is {\it a priori} unknown.

The BVP was solved numerically in \cite{SYT2} for the same class of scalar-tensor theories we consider in the present paper, and phases of
the black-hole solutions were found when $\beta<0$, i.e., in some regions of the parameter space, the number of black-hole solutions
with the same mass and charge may be more than one, so an additional parameter must be introduced in order to label the different solutions.
It is convenient to choose this
parameter to be the scalar charge\footnote{Another proper choice could be the value of scalar field on the horizon.
Let us also make some clarifying remarks on the non-uniqueness of the black-hole solutions.
It is twofold. For fixed values of the mass $M$ and charge $P$ the BVP described in this section has multiple solutions describing
black holes with different scalar charges but also different radii of the event horizon. From mathematical point of view it would
be unusual to interpret this situation as non-uniqueness since the different solutions have different domains $[r_H,\infty)$. The
presented formulation  of the problem is more germane to the theory of black holes since the classification of solutions is usually
based on the conserved charges such as the mass and electric or magnetic charge. However, another formulation of the problem is
possible where the more usual for mathematics non-uniqueness of solution is observed. If we fix the charge $P$ and the domain of
integration $[r_H,\infty)$ we obtain black-hole solutions which have the same radius of the event horizon but different masses $M$.} defined by
\begin{equation}
{\cal D} = - \lim_{r\rightarrow \infty} r^2 \frac{d\varphi}{dr}.
\end{equation}

\begin{figure}[htbp]%
\includegraphics[width=0.50\textwidth]{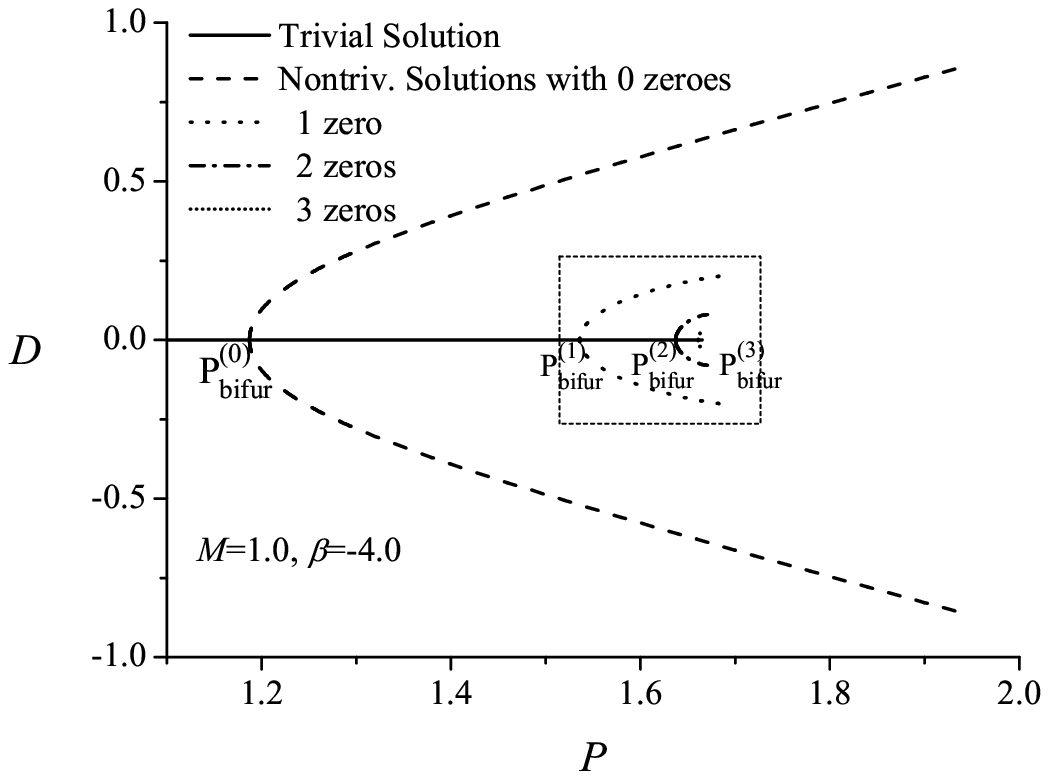}
\includegraphics[width=0.53\textwidth]{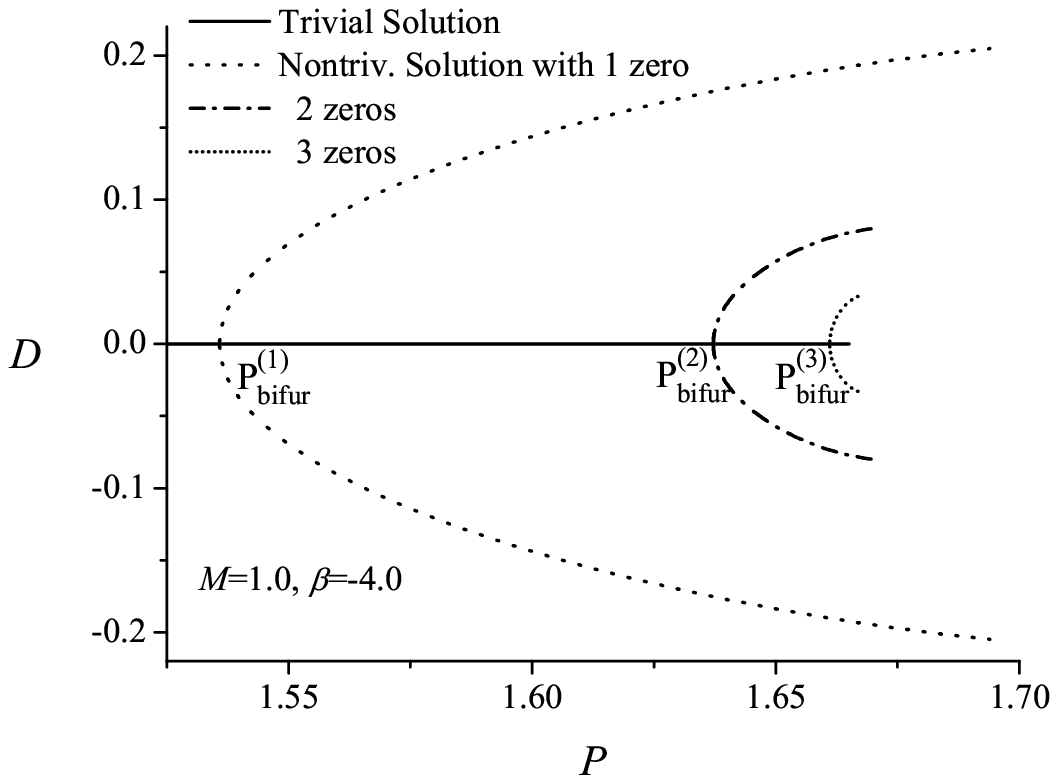}
\caption{The ${\cal D}(P)$ dependence for branches of black-hole solutions for fixed value of the mass $M=1.0$ and for $\beta=-4$.
The trivial branch of solutions and the nontrivial branches characterized
by scalar field which has up to three zeros are shown. The right panel is a magnification of the enclosed region of the
${\cal D}(P)$ diagram on the left panel.} \label{fig:D(P)M1beta-4}
\end{figure}%

\begin{figure}[htbp]%
\includegraphics[width=0.50\textwidth]{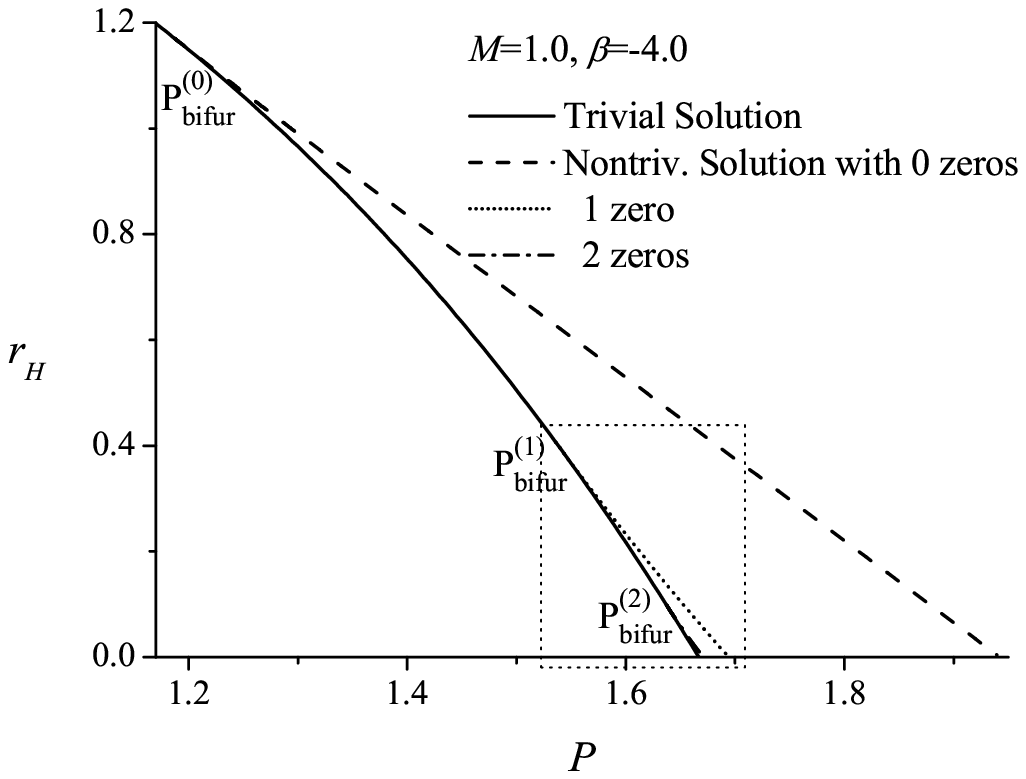}
\includegraphics[width=0.53\textwidth]{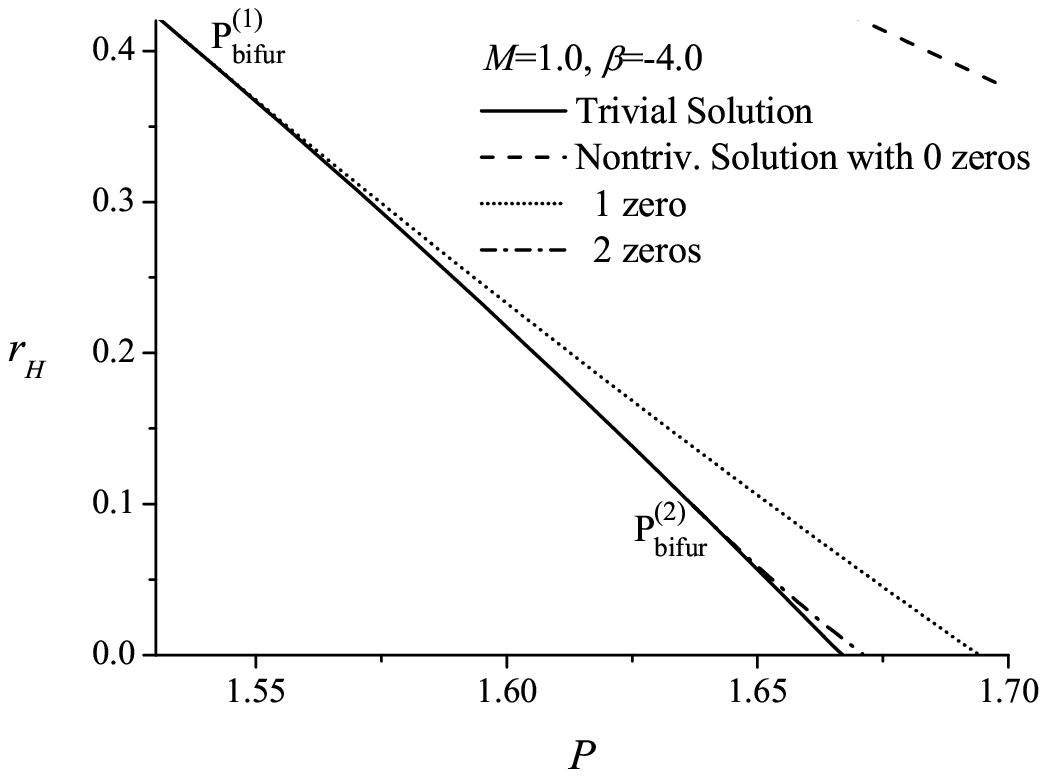}
\caption{The radius of the horizon $r_H$ as a function of the charge $P$ for some of the branches of black-hole solutions on Fig. \ref{fig:D(P)M1beta-4}.
The right panel is a magnification of the enclosed region of the $r_H(P)$ diagram on the left panel.} \label{fig:rh(P)M1beta-4}
\end{figure}%

Here, for the numerical solution of the above defined BVP, we apply two numerical methods in order to verify our results
-- the first method is the continuous analog of the Newtonian method that has been applied in \cite{SYT2} and the other
one is the shooting method \cite{NR}. The solutions obtained with these two methods are in agreement.

After performing a thorough search for new black-hole solutions, we indeed found that the values of the charge $P_\mathrm{crit}^{(k)}$ where
the unstable modes corresponding to the $k$-th bound state becomes zero and the values of the charge
$P_\mathrm{bifur}^{(k)}$ corresponding to the $k$-th bifurcation point on the phase diagram ${\cal D}(P)$ coincide. An example for this correspondence
can be seen on Figs. \ref{fig:BoundStateM1} and \ref{fig:D(P)M1beta-4} where the unstable modes and the branches of
black-hole solutions are shown for values of the parameters $M=1$ and $\beta=-4$. The charges $P_\mathrm{crit}^{(k)}$
and $P_\mathrm{bifur}^{(k)}$ are denoted on both figures, and for each value of $k$, they coincide within
$0.2\%$ error. On Fig. \ref{fig:rh(P)M1beta-4} the $r_H(P)$ dependence is shown for some of the branches of black-hole solutions presented on
Fig. \ref{fig:D(P)M1beta-4}.

\begin{figure}[htbp]%
\includegraphics[width=0.50\textwidth]{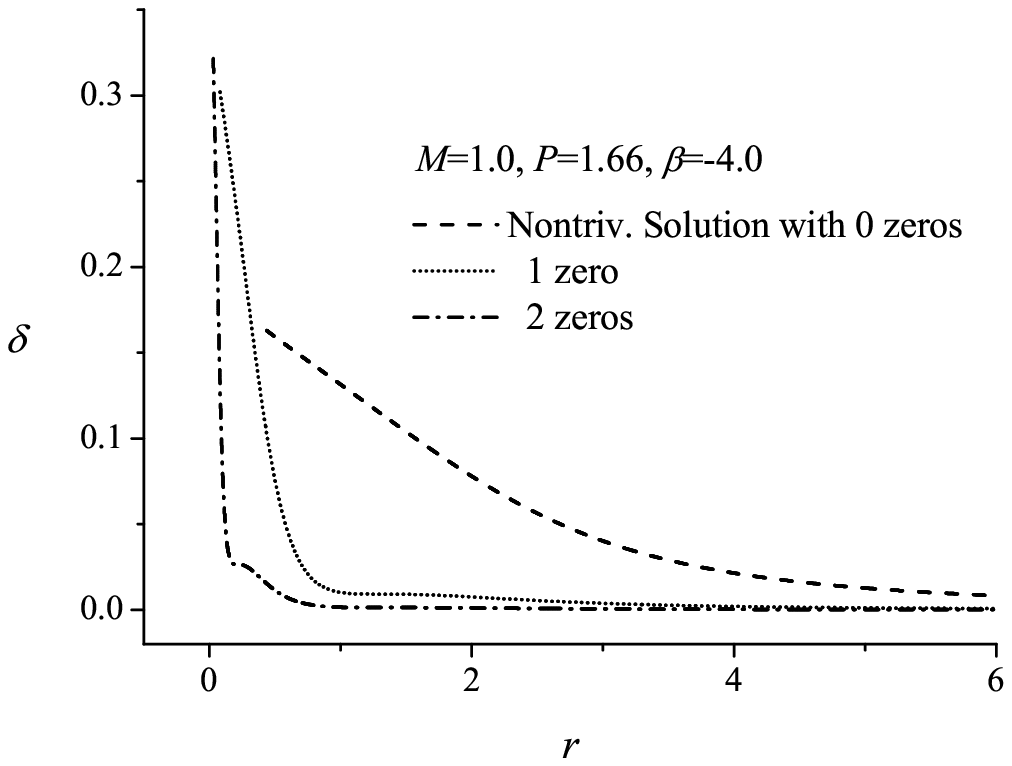}
\includegraphics[width=0.53\textwidth]{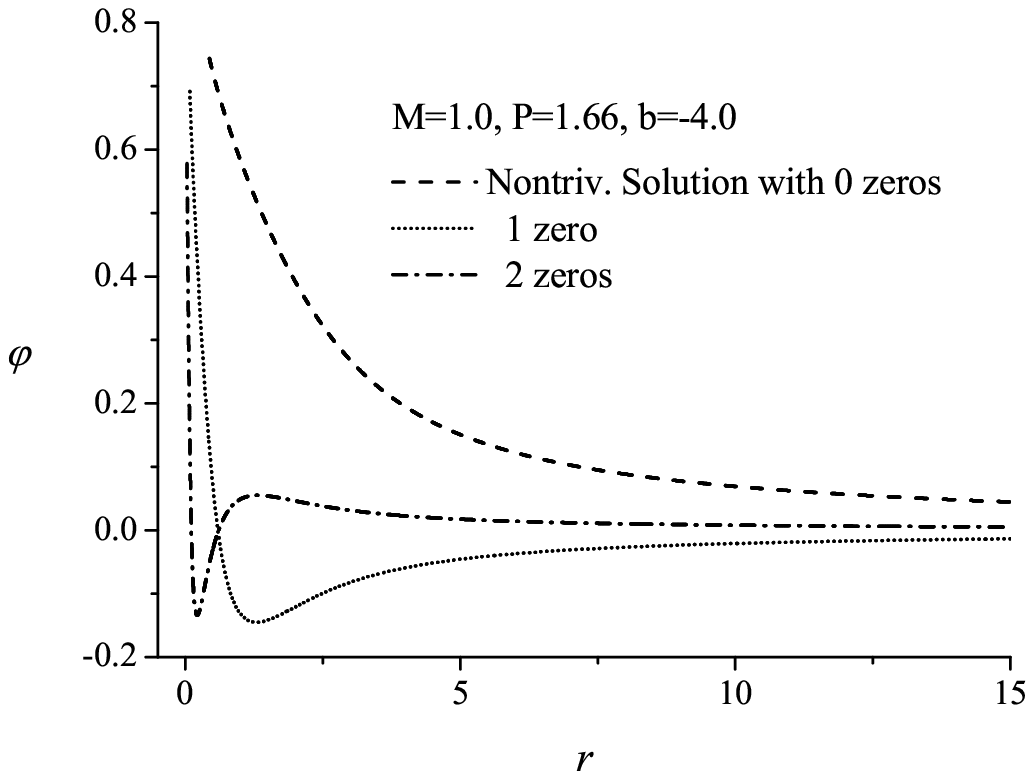}
\caption{The metric function $\delta(r)$ (left panel) and the scalar field $\varphi(r)$ (right panel) for nontrivial black-hole solutions
for $M=1.0$, $P=1.66$, and $\beta=-4$. The
solutions characterized by monotonous scalar field and by scalar field which has one and two zeros are shown.} \label{fig:delta_phi(r)M1beta-4}
\end{figure}

\begin{figure}[htbp]%
\includegraphics[width=0.50\textwidth]{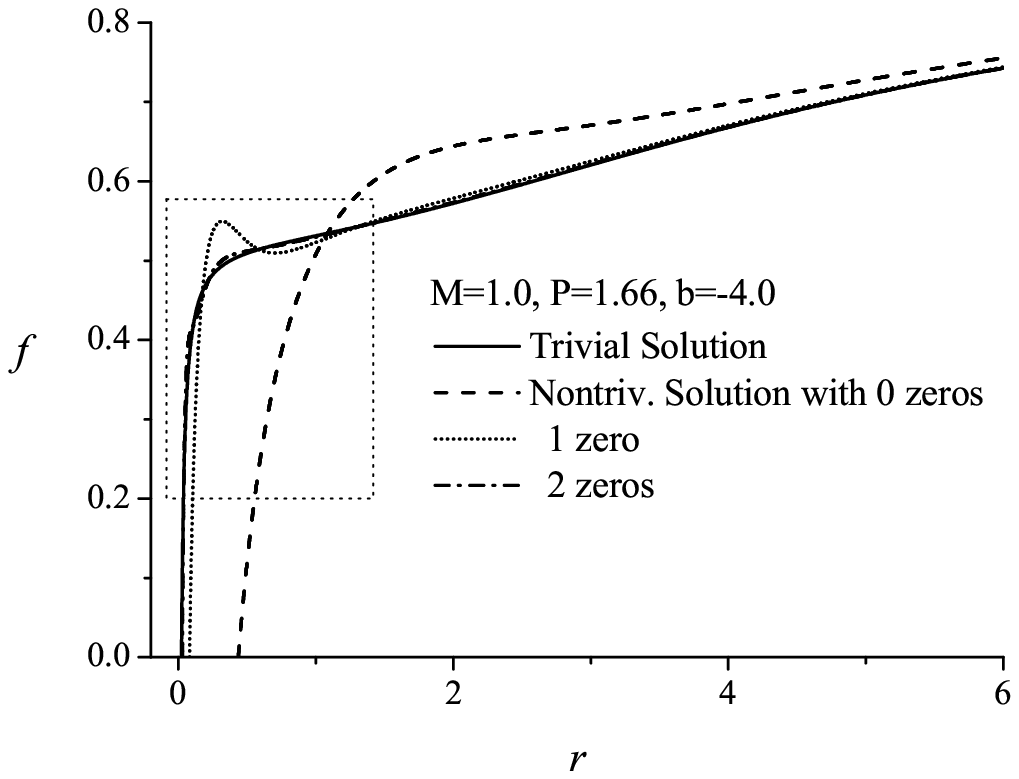}
\includegraphics[width=0.53\textwidth]{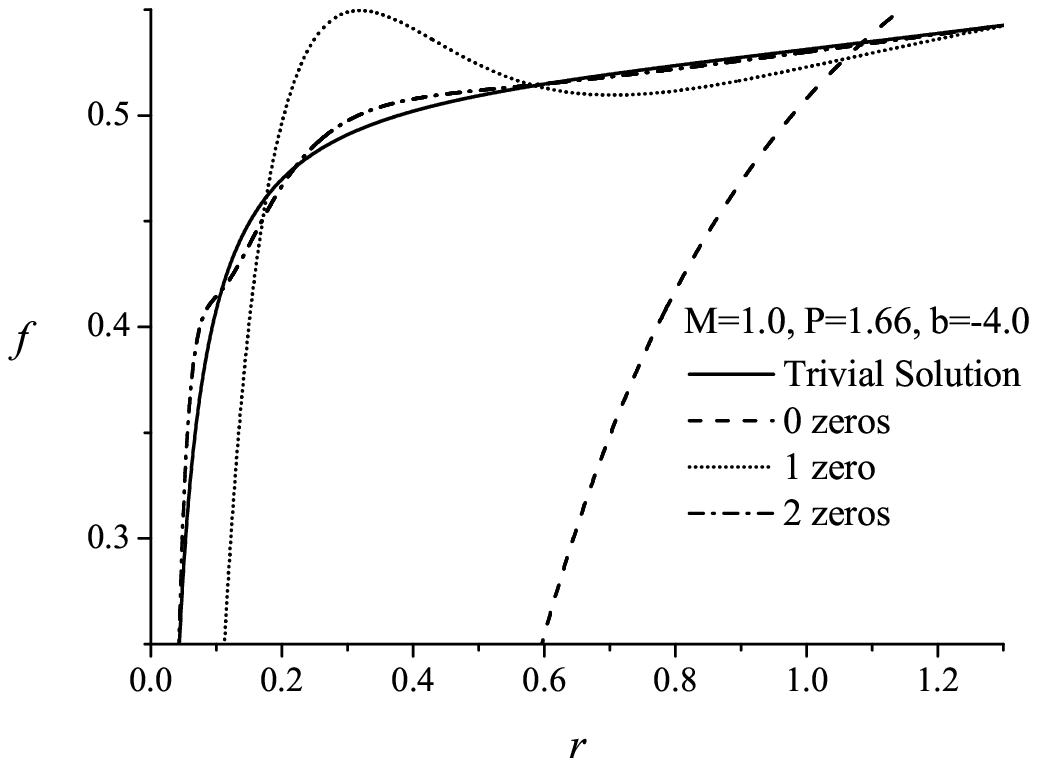}
\caption{The metric function $f(r)$ for the same black-hole solutions as on Fig. $\ref{fig:delta_phi(r)M1beta-4}$.
The right panel is a magnification of the enclosed region of the $f(r)$ dependence
on the left panel.} \label{fig:f(r)M1beta-4}
\end{figure}

As we said above, it is expected that the scalar field for the nontrivial solutions originating at
$P_\mathrm{bifur}^{(k)}$ has $k$ zeroes just as the perturbation function $\psi(r)$ for the unstable modes
corresponding to the $k$-th bound state. The metric functions $\delta(r)$ and $f(r)$ and the scalar field
$\varphi(r)$ of the nontrivial solutions for fixed values of the parameters $M=1$, $P=1.66$ and $\beta=-4$ are shown on
Figs. \ref{fig:delta_phi(r)M1beta-4} and \ref{fig:f(r)M1beta-4} and as we can see indeed the $k$-th black-hole solution is
characterized by a scalar field which has $k$ zeros. An interesting observation is that the function $f(r)$ for the nontrivial solutions
with scalar field which has one or more zeros  is not always monotonous  and
can have minima and maxima \footnote{It can be proven that the function $f(r)$ of the trivial solution
is always monotonous.}.

\section{Stability analysis of the nontrivial solutions}
The next step in our investigations is to determine whether the obtained new nontrivial black-hole solutions are stable
or not. Our approach is to study the QNMs associated with the pure radial perturbations of these solutions.
As we will not consider general perturbations, only the instability of a black-hole solution can be
proven rigorously within this approach. As  we will see below, the nontrivial solutions for which the scalar field has zeros, are indeed unstable.

The equations governing the radial perturbations of the nontrivial solutions were derived in \cite{SYT3}. It was shown in
\cite{SYT3} that the radial perturbations of the metric and the electromagnetic field  are determined by  the perturbations of the scalar field.
Hence, it is sufficient to consider only the perturbations of the scalar field $\delta\varphi$ which are governed by the equation

\begin{equation}\label{eq_phi_2}
\nabla_{\mu}^{(0)}{\nabla^{(0)}}^{\mu}\delta\varphi - U(r)\delta\varphi = 0,
\end{equation}
where
\begin{eqnarray}\label{eq:pot_nontriv}
&&U(r)=  -2 \Bigl[1+2r^2{\cal A}^{4}(\varphi)L\left(X^{(0)}\right)\Bigr]\Bigl[\partial_{r}\varphi(r)\Bigr]^2 \notag \\
&&\hskip 1.4cm +{\cal
A}^{4}(\varphi)\Bigl[16r\partial_{r}\varphi(r)\alpha(\varphi)+4\partial_{\varphi}\alpha(\varphi)\Bigr]
\Bigl[X^{(0)}\partial_{X^{(0)}}L\left(X^{(0)}\right) - L\left(X^{(0)}\right) \Bigr]  \\
&&\hskip 1.4cm -16\alpha^2(\varphi){\cal
A}^{4}(\varphi)\Bigl[\left(X^{(0)}\right)^{2}\partial^{2}_{X^{(0)}}L\left(X^{(0)}\right)-X^{(0)}\partial_{X^{(0)}}L\left(X^{(0)}\right)+L\left(X^{(0)}\right)\Bigr],
\notag
\end{eqnarray}
and the covariant derivative $\nabla_{\mu}^{(0)}$ and the function $X^{(0)}$ are with respect to the unperturbed background
solution. After separating the variables and introducing the tortoise coordinate $r_*$
\begin{equation}\label{subs}
\delta\varphi(r,t)=\frac{\psi(r)}{r} e^{i\omega t},\,\,\,\,\,\,\,\,\,\,\,\,\,\,\,\,\,dr_{*}={dr \over f(r)
e^{-\delta(r)}},
\end{equation}
we can obtain the following  Schr\"{o}dinger-like equation
\begin{equation}\label{eq:Schrod_NonTriv}
{d^{\,2} \psi(r_{*})\over dr_{*}^2}+\omega^2 \psi(r_{*})= U_{\rm {eff}}(r_{*})\psi(r_{*}),
\end{equation}
where the effective potential is given by
\begin{equation}\label{eq:Ueff_NonTriv}
U_{\rm {eff}}(r_{*})=f(r_{*}) e^{-2\delta(r_{*})}\Bigl\{U(r_{*})+2{\cal
A}^{4}(\varphi)L\left(X^{(0)}\right)+{1\over r^2(r_{*})}[1-f(r_{*})]\Bigr\}.
\end{equation}

\begin{figure}[htbp]%
\vbox{ \hfil \scalebox{0.80}{ {\includegraphics{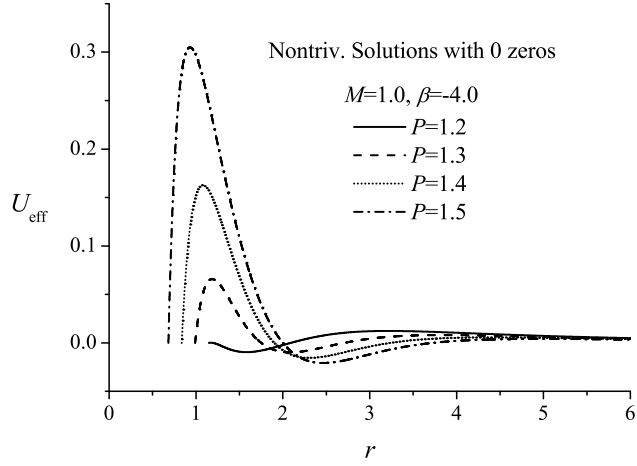}} }\hfil}%
\caption{The potential $U_{\rm eff}$ of the nontrivial solutions characterized by monotonous scalar field $\varphi(r)$ for $M=1$, $\beta=-4$ and
for several values of the magnetic charge $P$.} \label{fig:U(r)_nontriv_0zeros}
\end{figure}%

\begin{figure}[htbp]%
\includegraphics[width=0.50\textwidth]{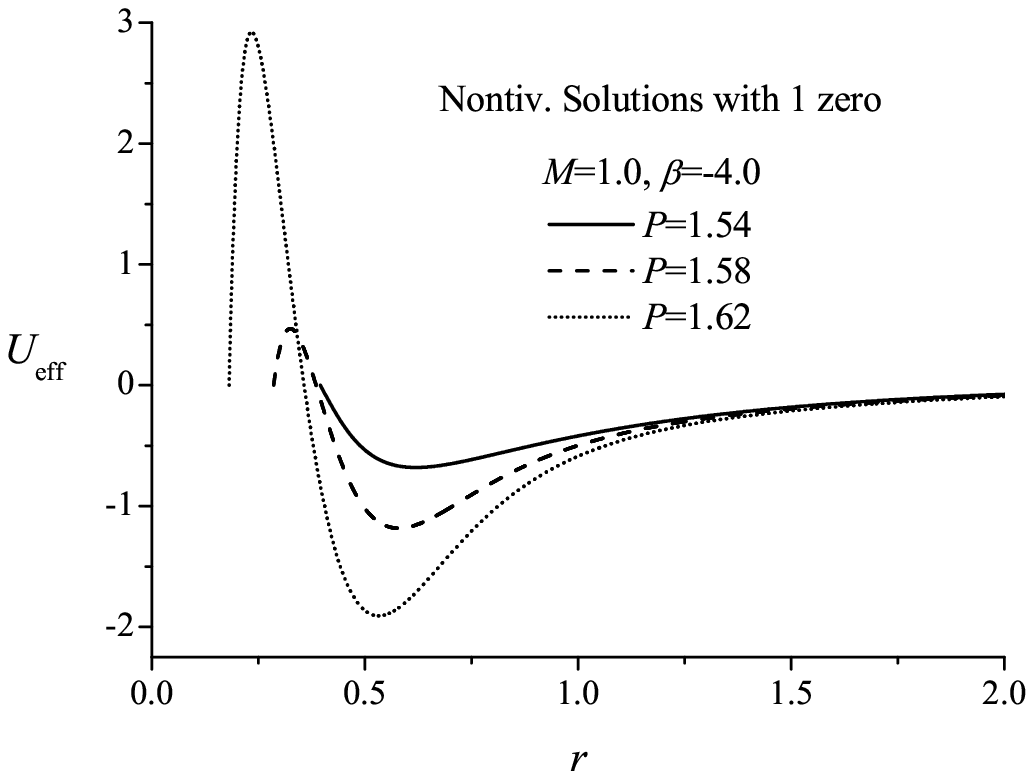}
\includegraphics[width=0.53\textwidth]{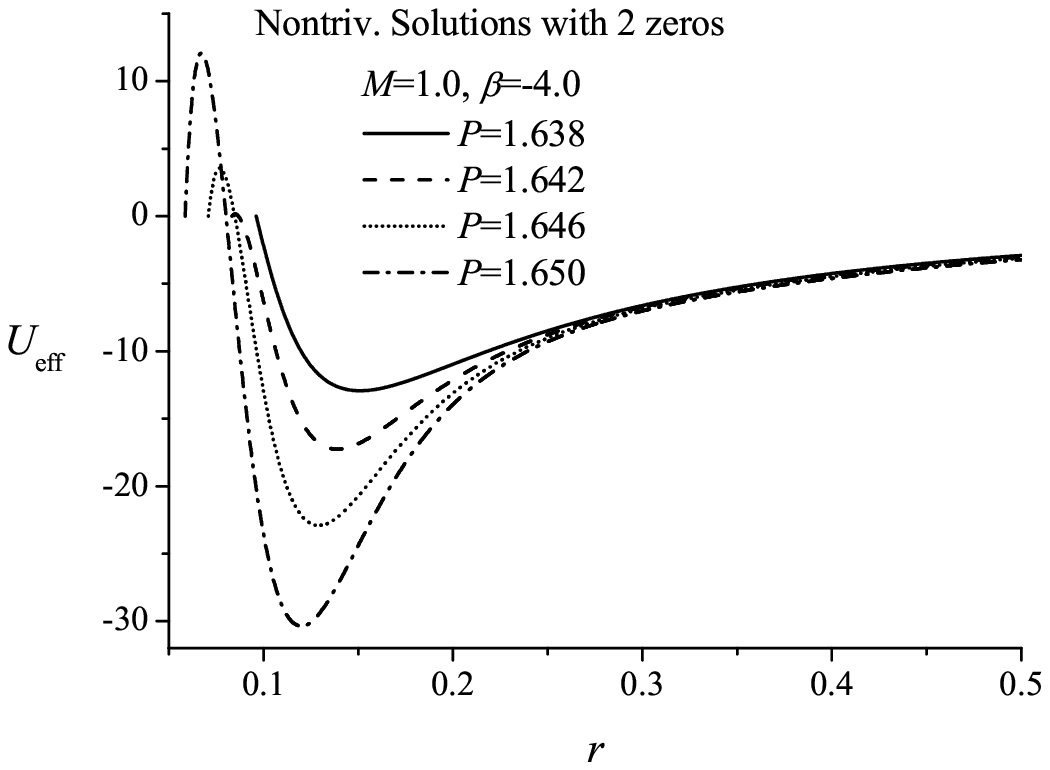}
\caption{The potential $U_{\rm eff}$ of the nontrivial solutions characterized by scalar field $\varphi(r)$ which has 1 zero (left panel) and
2 zeros (right panel) for $M=1$, $\beta=-4$ and
for several values of the magnetic charge $P$.} \label{fig:U(r)_nontriv_1_2zeros}
\end{figure}%

\begin{figure}[htbp]%
\vbox{ \hfil \scalebox{0.80}{ {\includegraphics{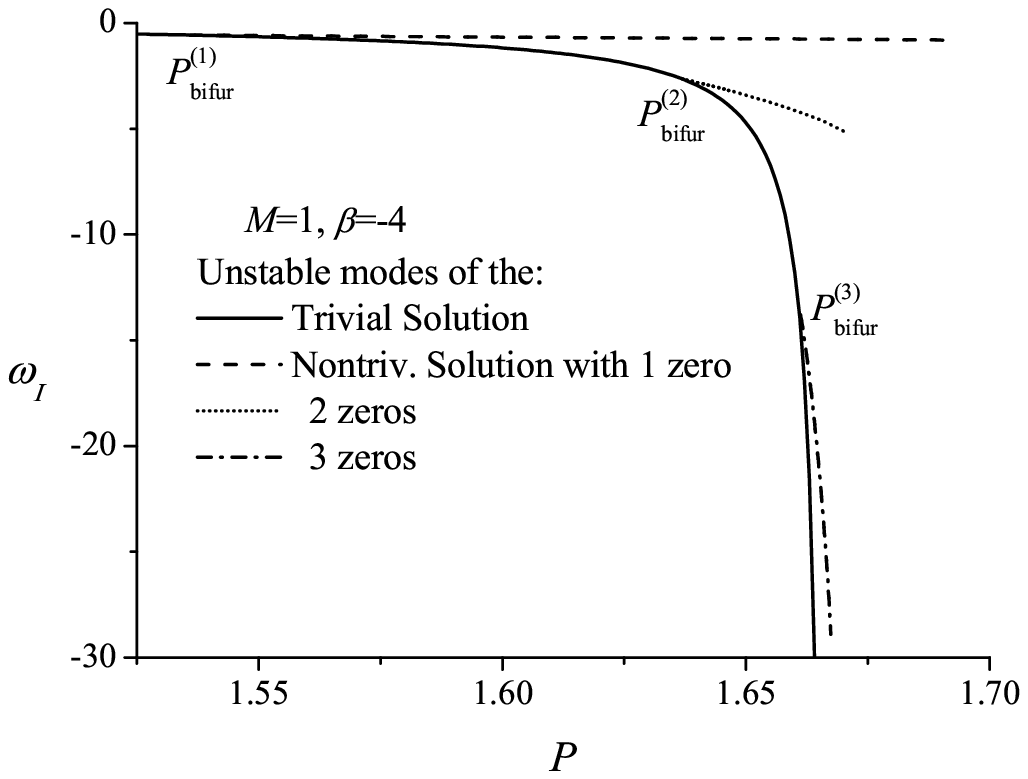}} }\hfil}%
\caption{The imaginary part of the unstable modes (the real part is zero) corresponding to the first bound state of the trivial solutions and
of the nontrivial solutions with scalar field $\varphi(r)$ which has up to three zeros ($M=1$ and $\beta=-4$).} \label{fig:EVNonTriv1_2zeros}
\end{figure}%

The effective potential of the nontrivial black-hole solutions has negative minimum for all of the studied values of the parameters which
means that unstable modes can exist. An example is given on Figs. \ref{fig:U(r)_nontriv_0zeros} and \ref{fig:U(r)_nontriv_1_2zeros} where the
effective potential $U_{\rm eff}(r)$ is plotted for the nontrivial solutions with scalar field which has no zeros and one or two
zeros.

We use the same numerical procedure for calculating the QNMs of the nontrivial solutions as the one described above.
The numerical results show
that for all of the studied values of the parameters the nontrivial solutions characterized by monotonous scalar field
are stable against radial perturbations. The new nontrivial solutions  characterized by scalar field
which has one or more zeros are unstable. The lowest unstable modes as a function of the charge are shown on Fig. \ref{fig:EVNonTriv1_2zeros}
for the trivial solution and for the nontrivial solutions characterized by scalar field which has up to three zeros
(the ${\cal D}(P)$ diagram of these black-hole solutions is given on Fig.~\ref{fig:D(P)M1beta-4}).
The branches of unstable modes of the nontrivial solutions bifurcate from the trivial one at some
critical charges which coincide with the charges $P^{(k)}_{\rm bifur}$ where
the nontrivial solutions bifurcate from the trivial one on the ${\cal D}(P)$ diagram  (Fig. \ref{fig:D(P)M1beta-4}). The results also show that for all
of the nontrivial solutions with $k$ zeros, $k$ bound states exist.

The results about the stability of the nontrivial solutions remain valid for all of the studied values of the
parameter. This means that all of the new nontrivial black hole branches characterized by
scalar field $\varphi(r)$ which has one or more zeros, are unstable, and the branches characterized by monotonous scalar
field are stable against radial perturbations. The results in the previous section show that the trivial solution
is stable against arbitrary\footnote{The trivial solution is stable against general linear perturbations of the metric and electromagnetic field
and this was demonstrated in \cite{MSarbach}.} (not only radial) perturbations before the first bifurcation point $P^{(0)}_{\rm crit}$.
We expect that after  the first bifurcation point $P^{(0)}_{\rm crit}$,  the solution possessing a scalar field without
zeros is  stable against arbitrary perturbations not only against the radial ones.

\section{Discussion}
In the present paper, we studied the scalar QNMs of the Einstein-Born-Infeld black-hole solution (the trivial solution) embedded  in a certain class
of scalar-tensor  theories with coupling function $\alpha(\varphi)=\beta\varphi$. The investigation of the spectrum of the QNMs shows that for $\beta<0$, there
are unstable modes which signal the presence of new nontrivial scalar-tensor, asymptotically flat, black-hole solutions with primary scalar hair that bifurcate from the trivial
solution. These solutions were constructed numerically by solving the full system of static and spherically symmetric field equations. The number of nontrivial solutions that bifurcate from the trivial one is equal to the number of the bound states of the
potential governing the scalar perturbations of the trivial solution. The nontrivial scalar-tensor black holes can be classified by the
number of the zeros of the scalar field in the black hole exterior. We have shown that the scalar-tensor black holes possessing scalar field
with one or more zeros are unstable.  It seems that only the scalar-tensor black holes having scalar field without any zeros are stable solutions.

The non-uniqueness of scalar-tensor black holes naturally raises the question of their classification. In other words, in the spirit
of the no-hair conjecture, the question is whether some kind of classification theorem (conjecture) could be
formulated. The most difficult question, however, is what kind of suitable further parameters associated with the black solutions
should be specified in addition to the conserved asymptotic charges (i.e., the mass and the charge). The situation in  some sense is similar to the situation in the
higher dimensions (see, for example, \cite{HY1},\cite{HY2}). One possible solution of the problem is the non-conserved scalar charge
${\cal D}$ to be included in the classification theorem as one of the necessary additional parameters just as the dipole charge in the uniqueness theorem for the higher dimensional Einstein-Maxwell gravity \cite{HY2}. Another obvious possibility is the inclusion of  some
"topological" characteristics of the scalar field (number of zeros, critical points)  as  additional parameters in the classification conjecture. Finally, we may adopt the view that the classical no-hair conjecture holds
but only for the stable solutions. This point of view, however, seems to be not true. Our preliminary investigations show that, though for
$\beta<-4.5$, probably  there exists a domain in the parameter space where the trivial solution and the solution possessing scalar field
without zeros are both linearly stable.  In any case, the raised questions will need much  deeper investigation.

The results in the present paper also open a way for studying dynamical processes in the strong field regime within the framework
of scalar-tensor theories. Especially, our results are the first steps to the study of strong field dynamical transitions from
unstable to stable scalar-tensor black hole configurations. On the level of presented perturbative analysis, the
characteristic  time scales of the transitions between the unstable and stable black holes are of order $\tau \sim 1/\omega$
where $\omega$ is the frequency of the corresponding  lowest unstable mode. A deeper investigation of the dynamics of
strong field transitions will require nonlinear analysis similar to that done in \cite{Salgado} for the scalar-tensor boson stars.

\section*{Acknowledgements}
This work was partially supported by the Bulgarian National Science Fund under Grants DO 02-257, VUF-201/06, by Sofia
University Research Fund under Grant No 101/2010. D.D. would like to thank the DAAD for a scholarship and the Institute
for Astronomy and Astrophysics T\"{u}bingen for its kind hospitality.

%%%%%%%%%%%%%%%%%%%%%%%%%%%%%%%%%%%%%%%%%%%%%%%%%%%%%%%%%%%%%%%%%%%%%%%%%%%%%%%


\begin{thebibliography}{9}
%%%%%%%%%%%%%%%%%%%%%%%%%%%%%%%%%%%%%%%%%%%%%%%%%%%%%%%%%%%%%%%%%%%%%%%%%%%%%%%

\bibitem{Kokkotas:1999bd}
 K.~D.~Kokkotas, B.~G.~Schmidt,
{\it Living Rev. Rel.}  {\bf 2}, 2 (1999).

\bibitem{Nollert:1999ji}
 H.~P.~Nollert,
{\it Class.\ Quant.\ Grav.\ } {\bf 16}, R159 (1999).
\bibitem{Dreyer} O. Dreyer, B. Kelly, B. Krishnan, L. S. Finn, D. Garrison, R. Lopez-Aleman, {\it Class. Quant. Grav.}~\textbf{21}, 787 (2004).

\bibitem{cardosotopical} E. Berti, V. Cardoso, A. Starinets,  {\it Class. Quant. Grav.} {\bf 26}, 163001 (2009).


\bibitem{RGCai1} J. Shen, B. Wang, C. Y. Lin, R. G. Cai, R. K. Su,   {\it JHEP}  {\bf 0707}, 037 (2007).
\bibitem{RaoWangYang} X. Rao, B. Wang, G. Yang, {\it Phys. Lett.} \textbf{B 649}, 472 (2007); arXiv:0712.0645 [gr-qc].
\bibitem{JingPan} J. Jing, Q. Pan, {\it Phys. Lett.} \textbf{B 660}, 13 (2008); arXiv:0802.0043 [gr-qc].
\bibitem{BertiCardoso_phase}E. Berti, V. Cardoso,  {\it Phys.Rev.} {\bf D 77}, 087501 (2008).
\bibitem{Papantopoulos} G. Koutsoumbas, E. Papantonopoulos, G. Siopsis,  {\it JHEP} {\bf 0805}, 107 (2008); arXiv:0801.4921[gr-qc].
\bibitem{RGCai2}X. He, B. Wang, R. G. Cai, C. Y. Lin, {\it Phys. Lett.} \textbf{B 688}, \textbf{Issue 2-3}, 230 (2010); arXiv:1002.2679~[gr-qc].


\bibitem{DEFPRL} T.~Damour~ and~G. Esposito-Farese, {\it Phys. Rev. Lett.} {\bf 70}, 2220 (1993).


\bibitem{Saa} A.~Saa,~{\it J.~Math.~Phys.}~{\bf 37},~2346~(1996).
\bibitem{BSen} N.~Banerjee~and~S.~Sen,~{\it Phys.~Rev.}~{\bf D58},~104024~(1998).
\bibitem{Hawking} S.W.~Hawking,~{\it Commun.~Math. Phys}~{\bf 25},~167~(1972).
\bibitem{Be1} J.~Bekenstein,~{\it Phys.~Rev.}~{\bf D5},~2403~(1972).
\bibitem{Be2} J.~Bekenstein,~{\it Phys.~Rev.~}{\bf D5},~1239~(1972).
\bibitem{Heusler} M.~Heusler, {\it Class.~Quant.~Grav.} {\bf 12},~2021~(1995).


\bibitem{SYT1}I.~Stefanov, S.~Yazadjiev~and M.~Todorov, {\it Phys.~Rev. }~{\bf D 75}, 084036 (2007).

\bibitem{SYT1a} I.~Stefanov, S.~Yazadjiev~and M.~Todorov, {\it Mod. Phys. Lett.}~{\bf A22}, 1217 (2007).
\bibitem{SYT2}I.~Stefanov,~S.~Yazadjiev~and~M.~Todorov, {\it Mod.~Phys.~Lett.} {\bf A23}, 2915 (2008).

\bibitem{DFBPE} T.~Damour~ and~G. Esposito-Farese, {\it Phys. Rev.} {\bf D 54} 1474 (1996).




\bibitem{Abramowitz} M. Abramowitz and I.A. Stegun, {\it  ``Handbook of Mathematical Functions''}, Dover Publications, 1972.

\bibitem{Fernando} S. Fernando and C. Holbrook, {\it Int. J. Theor. Phys.} {\bf 45}, 1630 (2006)

\bibitem{ChandraShooting} S. Chandrasekhar and S. Detweiler, {\it Proc. R. Soc. London}, \textbf{A 344}, 441 (1975).
\bibitem{KokkotasRN} K.D. Kokkotas and B.F. Schutz, {\it Phys. Rev.} \textbf{D 37}, 3378 (1988).

\bibitem{NR} W. H. Press, S. A. Teukolsky, W. T. Vetterling and B. P. Flannery, {\it ``Numerical Recipes''}, Cambridge University Press,
Cambridge, 2007


\bibitem{TornBook} K.S. Thorne, in {\it  ``Theoretical Principles in Astrophysics and Relativity''} ed. N.R. Lebowitz, W.H. Reid and P.O. Vandervoort,
(The University of Chicago Press, Chicago, 1978).
\bibitem{Harada} T. Harada, {\it Prog. Theor. Phys.} {\bf 98}, 359 (1997)



\bibitem{SYT3}I.~Stefanov,~S.~Yazadjiev~and~M.~Todorov, {\it Class. Quantum Grav.}  {\bf 26}, 015006 (2009).


\bibitem{MSarbach} C.~Moreno~and~O.~Sarbach,  {\it Phys. Rev.} {\bf D67}, 024028 (2003); [arXiv:gr-qc/0208090];

\bibitem{HY1} S.~Hollands and S.~Yazadjiev, {\it Commun. Math. Phys.} {\bf 283}, 749 (2008); [arXiv:0707.2775 [gr-qc]].

\bibitem{HY2}  S.~Hollands and S.~Yazadjiev,
  {\it Class.\ Quant.\ Grav.\ } {\bf 25}, 095010 (2008); [arXiv:0711.1722[gr-qc]]

\bibitem{Salgado} M. Alcubierre, J. Degollado, D. Nunez, M. Ruiz and M. Salgado,  {\it Phys. Rev.} \textbf{D 81}, 124018 (2010);arXiv:1003.4767v2 [gr-qc]

\end{thebibliography}
\end{document}